\documentclass[12pt,a4paper%,draft
]{amsart}
\usepackage{ifdraft}
\usepackage{darkmode}
\usepackage{amssymb,amsbsy}
\usepackage[obeyDraft]{todonotes}
\makeatletter\@mparswitchfalse\makeatother\normalmarginpar
\usepackage{bbm}
\usepackage[latin1]{inputenc}
\ifdraft{\textwidth145mm}{ \textwidth165mm}
\usepackage{enumerate}
\usepackage{mathtools}
\usepackage{subfig}

\usepackage{hyperref}
\hypersetup{
    colorlinks=true,
    linkcolor=blue,
    filecolor=magenta,
    urlcolor=cyan,
    pdftitle={Overleaf Example},
    pdfpagemode=FullScreen,
}

\newcommand{\bE}{\mathbb{E}}
\newcommand{\bF}{\mathbb{F}}
\newcommand{\bG}{\mathbb{G}}
\newcommand{\bH}{\mathbb{H}}
\newcommand{\bN}{\mathbb{N}}
\newcommand{\bR}{\mathbb{R}}

\newcommand{\cA}{\mathcal{A}}
\newcommand{\cD}{\mathcal{D}}
\newcommand{\cF}{\mathcal{F}}
\newcommand{\cG}{\mathcal{G}}
\newcommand{\cH}{\mathcal{H}}
\newcommand{\cJ}{\mathcal{J}}
\newcommand{\cS}{\mathcal{S}}
\newcommand{\cV}{\mathcal{V}}
\newcommand{\cX}{\mathcal{X}}

\newcommand{\sD}{\cD_\delta}
\newcommand{\supp}{{\rm supp}}
\newcommand{\dom}{{\rm Dom}}
\newcommand{\wprod}{\diamond}
\newcommand{\Normal}{\mathcal{N}}
\newcommand{\Var}{\mathbb{V}{\textrm ar}}

\theoremstyle{plain}
\newtheorem{theorem}{Theorem} [section]
\newtheorem{corollary}[theorem]{Corollary}
\newtheorem{lemma}[theorem]{Lemma}
\newtheorem{remark}[theorem]{Remark}
\newtheorem{proposition}[theorem]{Proposition}

\theoremstyle{definition}
\newtheorem{definition}[theorem]{Definition}

\numberwithin{figure}{section}

\begin{document}
    \title[Time evaluation of portfolio for asymmetrically informed traders]{Time evaluation of portfolio for asymmetrically informed traders}

    \author[B. D'Auria \& C. Escudero]{Bernardo D'Auria and Carlos Escudero}
    \address{Mathematics Department, University of Padua, Padua, Italy\newline\indent Departamento de Matem\'aticas Fundamentales, Universidad Nacional de Educaci\'on a Distancia, Madrid, Spain}
    \email{bernardo.dauria@unipd.it\\cescudero@mat.uned.es}

    \keywords{Portfolio optimization; asymmetrically informed traders; anticipating stochastic calculus; Malliavin calculus; white noise analysis.
    \\ \indent 2020 \textit{MSC: 60H05, 60H07, 60H10, 60H40, 91G10, 91G80}}

    \date{\today}

    \begin{abstract}
        We study the anticipating version of the classical portfolio optimization problem in a financial market with the presence of a trader who possesses privileged information about the future (insider information), but who is also subjected to a delay in the information flow about the market conditions; hence this trader possesses an asymmetric information with respect to the traditional one.
        We analyze it via the Russo-Vallois forward stochastic integral, i.~e. using anticipating stochastic calculus, along with a white noise approach.
        We explicitly compute the optimal portfolios that maximize the expected logarithmic utility assuming different classical financial models: Black-Scholes-Merton, Heston, Vasicek. Similar results hold for other well-known models, such as the Hull-White and the Cox-Ingersoll-Ross ones. Our comparison between the performance of the traditional trader and the insider, although only asymmetrically informed, reveals that the privileged information overcompensates the delay in all cases, provided only one information flow is delayed. However, when two information flows are delayed, a competition between future information and delay magnitude enters into play, implying that the best performance depends on the parameter values. This, in turn, allows us to value future information in terms of time, and not only utility.
    \end{abstract}
    \maketitle

    \todo{A last couple of revision rounds are still needed}

    \todo{to be potentially sent to JMAA journal} % Journal of Mathematical Analysis and Applications

    \section{Introduction}\label{sec:intro}
    The optimization of a portfolio is one of the most studied problems within the field of mathematical finance along with others such as option pricing, see for instance~\cite{karatzas1, markowitz, merton1}. Remarkably from a methodological viewpoint, these types of problems require rather developed mathematical techniques.
    This fact dates back, at least, to the thesis of Louis Bachelier, entitled \textit{Th\`eorie de la sp\`eculation}, see~\cite{bachelier}, were the motion of the stock prices is assimilated to diffusion processes. This was nothing but the starting point of a very fruitful collaboration between finance and mathematics, as can be seen, for example, in~\cite{bsm,noep,jeanblanc2009,merton2} and many other references. The family of mathematical tools commonly employed to assess financial problems include stochastic analysis, calculus of variations, partial differential equations, etc.

    In this work we focus on the portfolio optimization that can be performed by traders who are asymmetrically informed. This is related to the problem of insider trading, in which it is quite common to compare the performance of two investors, one of them assumed to possess more information on the evolution of the stock market than the other. This extra amount of information is the privileged or insider information. The mathematical formalization of this financial problem makes use of advanced stochastic techniques, such as Malliavin calculus, as can be found in many references in the literature, for instance in~\cite{bastons2018triple, bo, noep, do1, do2, elizalde2022chances, elizalde2022apo, escudero2018, jeanblanc2009, leon, pk}. However, herein we do not treat one of the investors as possessing privileged information over the other. Rather than that, both investors are considered asymmetrically informed. One is a traditional investor who knows present and historic prices of the stock market. The other one possesses future information on the stock market, which can be considered as insider information, however, at the same time, s/he only is informed of present and historic prices with a temporal delay. This creates a competition between time scales of future and delayed information, and unraveling this is the goal of the present work.

    To mathematically formalize the problem, consider a Black-Scholes market with two assets. The first one is risk-free
    \begin{subequations}
        \begin{align}
            \label{eq:cbsm1a}
            d X_0(t) = \,  & \rho(t) \, X_0(t) \, dt, \\ \label{eq:cbsm1b}
            X_0(0)   = \,  & 1,
        \end{align}
    \end{subequations}
    and can be considered a bank account or a bond.
    The other is risky
    \begin{subequations}
        \begin{align}
            \label{eq:cbsm2a}
            d X_1(t) = \,  & \mu(t) \, X_1(t) \, dt + \sigma(t) \, X_1(t) \, dB(t), \\ \label{eq:cbsm2b}
            X_1(0)  = \, & x,
        \end{align}
    \end{subequations}
    where $x>0$, such as a stock.
    For the time being, we consider $\rho(t)$, $\mu(t)$, and $\sigma(t)$ to be deterministic functions of time defined on the finite interval $[0,T]$, with $T>0$; moreover we assume them to be both positive and continuous.
    Financially, $\rho(t)$ encodes the interest rate of the bond, $\mu(t)$ is the expected return rate of the stock, $\sigma(t)$ is its volatility, and $T$ is the time horizon of the investment. Specifically for the volatility we will assume the condition $||1/\sigma(\cdot)||_{\infty}<\infty$, and we will refer to this condition in words as {\it the volatility is bounded away from zero}; it will be a useful hypothesis to prevent divergences since the volatility routinely appears in the denominators of our expressions.
    To pose the second equation (that is, problem~\eqref{eq:cbsm2a}-\eqref{eq:cbsm2b}), we introduce a \emph{white noise probability space}, $\left( \Omega, \cF, \bF, P \right)$, where $\Omega=\cS'(\mathbb{R}^2)$ is the space of tempered distributions on the plane, $\cF$ is the family of all Borel subsets of $\Omega$ equipped with the weak$^*$ topology.
    We suggest~\cite{noep} as a good reference on the white noise probability space and the corresponding theory.
    Our choice for the probability space is justified by the fact that it is the natural platform to work with the Malliavin and the Wick calculi.
    On this space, we define a two-dimensional Wiener process
    $(B(t,\omega),W(t,\omega))$ as the continuous version of $<\omega,\chi_{[0,t]}^{\otimes2}>$, with $\omega\in\Omega$, $t\in[0,T]$ and where $\chi_A$ denotes the indicator function of the set $A\subset[0,T]$.
    Let $\bF = \lbrace \cF_t \rbrace_{t\in [0,T]}$  be the natural filtration generated by the Brownian motion $B(t)$, i.e. $\cF_t = \lbrace \sigma(B(s)): 0 \leq s \leq t \rbrace$, and by $\bH = \lbrace \cH_t \rbrace_{t\in [0,T]}$, the corresponding one for $W(t)$.
    By construction, $B(t)$ and $W(t)$ are two independent Brownian motions.

    Now assume a trader wants to build a portfolio on this market: this is mathematically described by the control process $\pi \left(t\right)$ representing the fraction of the total wealth of this trader, hereafter denoted by $X^{\pi}(t)$, invested in the stock at time $t$. From now on we refer to $\pi(t)$ simply as the portfolio. If this portfolio is self-financing, then the total wealth process $X^{\pi}(t)$ at time $t\in[0,T]$ solves the stochastic differential equation
    \begin{subequations}
        \begin{align}
            \label{eq:cbsm3a}
            dX^{\pi}(t) = \,  & \left( 1 - \pi \left(t\right) \right) \rho(t) \, X^{\pi}(t) \, dt + \pi \left(t\right) \, X^{\pi}(t) \left( \mu(t) \, dt + \sigma(t) \, dB(t) \right), \\
            \label{eq:cbsm3b}
            X^{\pi}(0)  = \,  & x;
        \end{align}
    \end{subequations}
    note that this derivation holds both for the It\^o integral and for the forward integral (see Definition~\ref{rvfint} below) by virtue of Proposition 1.1 in~\cite{russovallois} and the assumptions on the coefficients (in symbols, the first and last $d$ in~\eqref{eq:cbsm3a} are interchangeable by $d^-$, see again Definition~\ref{rvfint}).
    Denote by $\cA$ the collection of all admissible self-financing portfolios. A portfolio $\pi(t)$ is considered to be admissible whenever it is an $\bF$-adapted stochastic process that is square-integrable, i.e.
    \begin{equation*}
        \bE \left[ \int_0^t \pi^2(s) \, ds \right] \, < \, \infty.
    \end{equation*}
    The assumption of $\bF$-adaptability means, at the financial level, that the trader can only build a portfolio based on the present and historic prices of the stock, but not the future ones. Such a trader is what we will call, from now on, a {\it traditional trader}.

    For any $\pi(t) \in \cA$, the stochastic differential equation~\eqref{eq:cbsm3a}-\eqref{eq:cbsm3b} fulfils the usual hypotheses of the existence and uniqueness theorem for It\^o stochastic differential equations~\cite{kuo,oksendal1}; therefore it follows that there exists a unique strong solution $X^{\pi}(t)$ in $t \in [0,T]$ for any fixed $T>0$. Moreover, this solution admits an explicit representation formula, which is readily computable by means of It\^o calculus; it reads
    \begin{align*}
        X^{\pi}(t) = \,  & x \, \exp \bigg \lbrace \int_0^t \sigma(s) \, \pi(s) \, dB(s)                                                                        \\
        & + \int_0^t \left(\rho(s) + \left(\mu(s) - \rho(s) \right) \pi(s) - \frac{1}{2} \, \sigma^2(s) \, \pi^2(s) \right) \, ds \bigg \rbrace.
    \end{align*}
    From now on we refer to the optimal portfolio $\bar{\pi}(t)$ as the control process $\pi(t)$ that maximizes the expectation of the logarithm of the trader wealth at time $T$, i.e. the admissible process that maximizes the quantity $\bE [\log (X^{\pi}(T))]$. In other words, the trader preferences are encoded by the theory of utilities, and precisely these traders are risk-averse and seek for the maximization of their logarithmic utility. This assumption follows both from financial modelling, as risk-aversion is common among traders, and mathematical convenience, since it is well known that the logarithmic utility favors the computation of explicit solutions. The problem we have just stated is classical~\cite{karatzas1, markowitz, merton1}, and the optimal portfolio is known to have the following explicit form
    \begin{equation}
        \label{eq:merton.pi}
        \bar\pi(t) = \frac{\mu(t) - \rho(t)}{\sigma^2(t)}.
    \end{equation}
    Hereafter we use this result as a benchmark, and we devote Section~\ref{sec:bsm} to compare it to the corresponding result that can be obtained for the {\it asymmetrically informed trader (AIT)}, that is, the one who possesses insider information about the future, but only partial information about the past.

    The rest of this paper is organized as a follows. In Section~\ref{sec:fm} we introduce some models that are quite popular in the mathematical finance literature. Section~\ref{sec:rvmc} is devoted to introduce the mathematical machinery (such as the forward integral) that is needed from then onwards. In Section~\ref{sec:bsm}, we formulate the portfolio optimization problem for asymmetrically informed traders in the case of a Black-Scholes market. Next, in Sections~\ref{sec:hes},~\ref{sec:vk}, and~\ref{sec:vk.delay} we analyze this problem for the different financial models that were previously introduced. Finally, in Section~\ref{sec:end}, we draw our main conclusions.

    \section{Financial models}\label{sec:fm}

    In this section we briefly introduce some of the financial models we are going to consider next. First of all, we note that the continuous compounding and Black-Scholes models were already introduced in equations~\eqref{eq:cbsm1a}-\eqref{eq:cbsm1b} and~\eqref{eq:cbsm2a}-\eqref{eq:cbsm2b} respectively, where the financial parameters were assumed to be deterministic continuous functions. The other models can be thought of as refinements of these two, which are the simplest ones. They follow from promoting some of those deterministic financial parameters to stochastic processes.

    The Heston model~\cite{hull, heston} assumes a stochastic volatility for the stock. That is, the volatility of the geometric Brownian motion modeling the stock price in equation~\eqref{eq:cbsm2a}, instead of being a deterministic $\sigma(t)$, becomes a stochastic process $V(t)$, which is given by the solution of the stochastic differential equation
    \begin{subequations}
        \begin{align}
            \label{eq:hes1a}
            dV(t) = \,  & \kappa \, (\theta - V(t)) \, dt + \eta \, \sqrt{V(t)} \, dW(t),
            \\ \label{eq:hes1b}
            V(0)  = \,  & v_0,
        \end{align}
    \end{subequations}
    where $\kappa$, $\theta$, $\eta$, and $v_0$ are positive constants. Financially, $\kappa$ is the mean reversion rate, $\theta$ is the asymptotic mean level of the volatility, and $\eta$ encodes the amplitude of the fluctuations.

    For the Vasicek model~\cite{hull, lambertonlapeyre, vasicek} what fluctuates is the interest rate rather than the volatility. This means that the drift of the geometric Brownian motion in~\eqref{eq:cbsm2a} is promoted from a deterministic function $\mu(t)$ to a stochastic process $R(t)$. In particular, the interest rate $R(t)$ is assumed to be given by the solution to the equation
    \begin{subequations}
        \begin{align}
            \label{eq:vk1a}
            dR(t) = \,  & a \, (b - R(t)) \, dt + \xi \, dW(t),
            \\ \label{eq:vk1b}
            R(0)  = \,  & r_0,
        \end{align}
    \end{subequations}
    where $b$, $a$, $\xi$, and $r_0$ are positive real numbers. Financially, $b$ is the asymptotic mean level for the interest rate, $a$ is the mean reversion rate, and $\xi$ is the interest rate diffusion.
    Obviously, the stochastic process $R(t)$ is nothing but an Ornstein-Uhlenbeck process, and therefore
    \begin{align*}
        R(t) &= r_0 \, e^{-at} + b \left(1-e^{-at}\right) + \xi \, e ^{-at} \int_0^t e^{as} \, dW(s), \\
        \bE[R(t)] &= r_0 \, e^{-at} + b \left(1-e^{-at}\right), \\
        \bE[R(t)^2] &= b^2 + \frac{\xi^2}{2 a}(1-e^{-2at}) + 2 \, b \, e^{-at}(r_0-b) + e^{-2at}(r_0 - b)^2.
    \end{align*}

    Another model for the short-rate dynamics is the one-factor Hull-White (HW) model~\cite{hull,hullwhite}. In this case $R(t)$ solves
    \begin{subequations}
        \begin{align}
            \label{eq:hw1a}
            dR(t) = \,  & \left(\kappa(t) - a \, R(t)\right) dt + \theta \, dW(t), \\
            \label{eq:hw1b}
            R(0)  = \,  & r_0.
        \end{align}
    \end{subequations}
    This model extends the previous one in the sense that $\kappa=\kappa(t)$ is no longer a constant but a function of time. In the present context it is enough to assume $\kappa(t)$ to be a deterministic function that is both continuous and positive in the interval $[0,T]$. As in the previous case, the stochastic process $R(t)$ is an Ornstein-Uhlenbeck, and therefore a Gaussian, process; in consequence, the explicit representation formulas for the process itself as well as its moments are readily computable.

    Finally, the Cox-Ingersoll-Roll (CIR) model~\cite{hull, lambertonlapeyre, cir} for the short rate is given by the equation
    \begin{subequations}
        \begin{align}
            \label{eq:cir1a}
            dR(t) = \,  & a \, \left(b - R(t)\right) dt + \theta \, \sqrt{R(t)} \, dW(t), \\ \label{eq:cir1b}
            R(0)  = \,  & r_0;
        \end{align}
    \end{subequations}
    herein $a$, $b$, $\theta$, and $r_0$ are again positive constants. Their financial meaning is the same as the one specified for the other models.

    It is evident that models~\eqref{eq:hes1a}-\eqref{eq:hes1b} and~\eqref{eq:cir1a}-\eqref{eq:cir1b} are formally equivalent although they model different things. The following proposition states some of the properties of the solution to this stochastic differential equation that will be useful in the remainder of this work.

    \begin{proposition}
        \label{prop:cir}
        Let $T>0$ be fixed but otherwise arbitrary.
        The stochastic differential equation
        \begin{subequations}
            \begin{align}
                dZ(t) = \,  & \kappa \, (\theta - Z(t)) \, dt + \eta \, \sqrt{Z(t)} \, dW(t), \\
                Z(0)  = \,  & z_0,
            \end{align}
        \end{subequations}
        possesses a unique strong solution in the interval $[0,T]$ for any positive constants $z_0, \kappa, \theta, \eta$.
        The solution stays positive almost surely if and only if $\kappa \theta \ge \eta^2/2$ (the so-called Feller condition).

        Furthermore, $\bE[Z(t)^{-1}] <\infty$ for every $t \in [0,T]$ and any $T>0$, if and only if $\kappa \theta \ge \eta^2$.
    \end{proposition}

    \begin{proof}
        The existence and uniqueness of solution follows directly from a theorem by Yamada and Watanabe, namely Theorem 1 in~\cite{watyam}. Moreover, the stochastic process $Z(t)$ stays positive for every $t \in [0,T]$, since the initial condition is positive and it can be explicitly represented with the formula
        \begin{equation*}
            Z(t)=\exp(-\kappa t) \, \text{BESQ}^\delta \left(\eta^2[\exp(\kappa t)-1]/(4 \kappa)\right),
        \end{equation*}
        where $\text{BESQ}^\delta(t)$ is a squared Bessel process with dimension $\delta = 4 \kappa \theta / \eta^2 \ge 2$ and initialized at $z_0$, see Chapter 6 in~\cite{jeanblanc2009}. On the other hand, if $\kappa \theta < \eta^2/2$, then $\text{BESQ}^\delta(t)$ becomes a squared Bessel process of dimension $\delta = 4 \kappa \theta / \eta^2 < 2$, and therefore positivity no longer holds, see Chapter 6 in~\cite{jeanblanc2009} again. Obviously, in such a case, $\bE[Z(t)^{-1}] <\infty$ does not hold either.

        From now on and as a consequence of the previous paragraph, we assume $\kappa \theta \ge \eta^2/2$. Since $f(z) = 1/z$ is a smooth function over the half-line $]0,\infty[$ and $Z(t)$ stays positive, one may use It\^o formula~\cite{kuo,oksendal1} to find
        \begin{equation*}
            d \left[ \frac{1}{Z(t)} \right]
            = \left(\frac{-\kappa\left(\theta - Z(s)\right)}{Z^2(s)} + \frac{\eta^2}{Z^2(s)} \right) ds - \eta \, \frac{1}{\sqrt{Z^3(t)}} \, dW(t).
        \end{equation*}
        Then, by the linearity of the expectation
        \begin{align*}
            \bE \left[ \frac{1}{Z(t)} \right]
            & = \bE \left[ \frac{1}{Z(0)} \right] + \bE \left[ \int_0^t \left(\frac{-\kappa\left(\theta - Z(s)\right)}{Z^2(s)} + \frac{\eta^2}{Z^2(s)} \right) ds \right]                                    \\
            & = \frac{1}{z_0} + \kappa \, \bE \left[ \int_0^t \frac{1}{Z(s)} \, ds \right] + \left(\eta^2 - \kappa \, \theta \right) \bE \left[ \int_0^t \frac{1}{Z^2(s)} \, ds \right].
        \end{align*}
        Interchanging the order of integration results in
        \begin{equation*}
            \bE \left[ \frac{1}{Z(t)} \right]
            = \frac{1}{z_0} + \kappa \int_0^t \bE\left[ \frac{1}{Z(s)} \right] ds + \left(\eta^2 - \kappa \, \theta \right) \int_0^t \bE\left[ \frac{1}{Z^2(s)} \right] ds;
        \end{equation*}
        or alternatively
        \begin{equation*}
            \frac{d}{dt} \, \bE \left[ \frac{1}{Z(t)} \right]
            = \kappa \, \bE\left[ \frac{1}{Z(t)} \right] + \left(\eta^2 - \kappa \, \theta \right) \bE\left[ \frac{1}{Z^2(t)} \right],
        \end{equation*}
        with $\bE \left[ 1/Z(0)\right]
        = 1/z_0$.
        Thus, if $\kappa \theta \geq \eta^2$:
        \begin{equation*}
            \frac{d}{dt} \, \bE \left[ \frac{1}{Z(t)} \right]
            \le \kappa \, \bE\left[ \frac{1}{Z(t)} \right],
        \end{equation*}
        so the Gr\"onwall inequality implies
        \begin{equation*}
            \bE \left[ \frac{1}{Z(t)} \right]
            \, \leq \, \frac{e^{\kappa t}}{z_0} \,
            \leq \, \frac{e^{\kappa T}}{z_0}.
        \end{equation*}
        On the contrary, if $\kappa \theta < \eta^2$:
        \begin{equation*}
            \frac{d}{dt} \, \bE \left[ \frac{1}{Z(t)} \right]
            \ge \left(\eta^2 - \kappa \, \theta \right) \bE\left[ \frac{1}{Z^2(t)} \right],
        \end{equation*}
        and by the Jensen inequality
        \begin{equation*}
            \frac{d}{dt} \, \bE \left[ \frac{1}{Z(t)} \right]
            \ge \left(\eta^2 - \kappa \, \theta \right) \bE^2\left[ \frac{1}{Z(t)} \right],
        \end{equation*}
        thus we conclude
        \begin{equation*}
            \bE \left[ \frac{1}{Z(t)} \right]
            \, \ge \, \frac{1}{z_0-\left(\eta^2 - \kappa \, \theta \right)t},
        \end{equation*}
        which implies a divergence as $t\to z_0/\left(\eta^2 - \kappa \, \theta \right)$.
    \end{proof}

    \begin{remark}
        The first two moments of the stochastic process $Z(t)$ are readily computable from the zero mean property of the It\^o integral and the It\^o isometry, and read
        \begin{align*}
            \bE[Z(t)] &= z_0 \, e^{-\kappa t} + \theta \left(1-e^{-\kappa t}\right),
            \\ \nonumber
            \bE[Z(t)^2] &= \left( z_0 \, e^{-\kappa t} + \theta \left(1-e^{-\kappa t}\right) \right)^2 + \frac{ z_0 \, \eta^2}{\kappa} \left(e^{-\kappa t} - e^{-2 \kappa t}\right) + \frac{\theta \, \eta^2}{2 \, \kappa} \left(1-e^{-\kappa t} \right).
        \end{align*}
    \end{remark}

From now on we will always assume the Feller condition for both models~\eqref{eq:hes1a}-\eqref{eq:hes1b} and~\eqref{eq:cir1a}-\eqref{eq:cir1b}, so their solutions always stay positive.

    \section{The Russo-Vallois forward stochastic integral and Malliavin Calculus}\label{sec:rvmc}
    
    To study the problems considered in this work, we analyze them via the anticipating stochastic calculus and using a white noise approach.
    Hence, in this section, we introduce the definition of the Russo-Vallois forward stochastic integral and some notions of Malliavin calculus, such as the Malliavin derivative and the Donsker delta function of a random variable.

    The Russo-Vallois forward stochastic integral~\cite{russovallois}, which was introduced by Francesco Russo and Pierre Vallois in 1993, generalizes the It\^o one~\cite{ito1, ito2} to anticipating integrands.
    Under suitable assumptions, it preserves It\^o calculus~\cite{russovallois,noep}, producing the same results as the latter when the integrand is adapted.
    
    \begin{definition}\label{rvfint}
        A stochastic process $\varphi(t)$ is \textit{forward integrable in the strong sense} with respect to the Brownian motion~$B(t)$ if there exists a stochastic process~$I(t)$ such that
        \begin{equation*}
            \sup_{t\in[0,T]} \left | \int_0^t \varphi(s) \, \frac{B(s+\varepsilon)-B(s)}{\varepsilon} \, ds - I(t) \right| \to 0, \ \ \ \ \mbox{as} \ \ \varepsilon \to 0^+,
        \end{equation*}
        in $L^2([0,T])$.
        In this case, $I(t)$ is the \textit{forward integral} of $\varphi(t)$ with respect to $B(t)$ on $[0,T]$, and we denote
        \begin{equation*}
            I(t) \coloneqq \int_0^t \varphi(s) \, d^- B(s),  \ \ \ \ t \in [0,T].
        \end{equation*}
    \end{definition}
    
    Now, in order to introduce the Malliavin derivative, we present the concept of chaos expansions in the Hida distribution space $(\cS)^*$, the dual of the space $(\cS)$.

    Following the notations of section~2 of~\cite{noep}, we denote by $\cJ$ the set of all finite multi-indices $\alpha=(\alpha_1, \alpha_2, \ldots, \alpha_m)$, with $m\in\bN$. We define $\alpha! \coloneqq \alpha_1!\cdots\alpha_m!$, $(2\bN)^\alpha \coloneqq (2)^{\alpha_1} \cdots (2m)^{\alpha_m}$ and the functions~$\{H_\alpha\}_{\alpha\in\cJ}$  as
    \begin{equation}
        \label{eq:H.fun}
        H_\alpha(\omega)
        \coloneqq
        \prod_{j=1}^m h_{\alpha_j}(\omega_{e_j}), \quad \omega\in\Omega.
    \end{equation}

    In the formula above, for $n\ge0$,  $h_n(x)$ denotes the $n$-th Hermite polynomial, defined as $h_n(x)\coloneqq \bE[(x + i Z)^n]$ with $Z$ a standard normal random variable.
    The random variable $\omega_\phi$ denotes the \emph{smoothed white noise}, defined as
    \begin{equation}
        \label{smooth.wn}
        \omega_\phi = \int_0^T \phi_t \, dB_t(\omega) ,
    \end{equation}
    where $\phi \in L^2([0,T])$ is a deterministic function, see (5.4) in~\cite{noep}.
    In particular, in~\eqref{eq:H.fun}, the function $e_j(x)$ is the so called Hermite function, defined as
    \begin{equation*}
        e_j(x) \coloneqq
        e^{-x^2/2}
        h_{j-1}(x\sqrt{2})/\sqrt{(j-1)!\sqrt{\pi}}.
    \end{equation*}
    The family of functions $\{H_\alpha\}_{\alpha\in\cJ}$
    constitutes an orthonormal basis of $L^2\left( \Omega, \cF, P \right)$, see Theorem~2.2.4 in~\cite{HOUZ10}.

    We now introduce the space of smooth random variables $(\cS)$.
    \begin{definition}
        We say that $F\in(\cS)$ if it admits the following form
        \begin{equation*}
            F = \sum_{\alpha\in\cJ} a_\alpha H_\alpha,
        \end{equation*}
        with $a_\alpha\in\bR$ such that
        $\sum_{\alpha\in\cJ} \alpha! a_\alpha^2 (2\bN)^{\alpha k} < \infty$
        for any $k>0$.
    \end{definition}

    The dual space, $(\cS)^*$, is the Hida distribution space whose members are defined in the following way.
    \begin{definition}
        We say that $F\in(\cS)^*$ whenever it admits the following form
        \begin{equation*}
            F = \sum_{\alpha\in\cJ} a_\alpha H_\alpha,
        \end{equation*}
        with $a_\alpha\in\bR$ such that
        $\sum_{\alpha\in\cJ} \alpha! a_\alpha^2 (2\bN)^{-\alpha k} < \infty$
        for a given $k>0$.
    \end{definition}

    We also define the Wick product of two $(\cS)^*$ random variables, which is the natural product in the Hida distribution space.

    \begin{definition}
        For any two given two Hida distributions,  $F = \sum_{\alpha\in\cJ} a_\alpha H_\alpha\in(\cS)^*$  and $G = \sum_{\alpha\in\cJ} b_\alpha H_\alpha\in(\cS)^*$ , their Wick product, $F \wprod G$, is defined as
        \begin{equation*}
            F \wprod G
            = \sum_{\alpha,\beta\in\cJ} a_\alpha b_\beta H_{\alpha+\beta}
            = \sum_{\gamma\in\cJ} \big(\sum_{\alpha+\beta = \gamma} a_\alpha b_\beta \; \big) H_{\gamma}.
        \end{equation*}
    \end{definition}

    We now define the Malliavin derivative, see also~\cite[Definition~6.5]{noep}.

    We set $\epsilon^{(k)} \coloneqq (0,0,...,1,0,...,0)$, as the infinite-size vector with only the $k$-th~component equals to one.

    \begin{definition}
        If $F = \sum_{\alpha\in\cJ} a_\alpha H_\alpha \in(\cS)^*$, we define the \emph{Malliavin derivative} of $F$ at $t$ in $(\cS)^*$, $D_t F$, as the random variable with the following expansion
        \begin{equation*}
            D_t F \coloneqq   \sum_{\alpha\in\cJ} \sum_{k=1}^{\infty} a_\alpha \alpha_k e_k(t) H_{\alpha-\epsilon^{(k)}},
        \end{equation*}
        whenever this sum converges in $(\cS)^*$.
        We denote $\dom(D_t)$ the set of all $F \in (\cS)^*$ for which the above series converges in $(\cS)^*$.
    \end{definition}

    The expectation of the forward integral is entangled with the Malliavin derivative, as expressed by the following result, see ~\cite[Corollary~8.19]{noep} and~\cite[Proposition~1]{oksendal2017}. 

    \begin{proposition}
        \label{rvmccorollary}
        Let $\varphi(t)$ be a c\'agl\'ad process, forward integrable in the strong sense, such that the limit $D_{t^+}\varphi(t)\coloneqq \lim_{s \to t^+} D_{s}\varphi(t)$ exists with convergence in $L^2([0,T] \times \Omega)$.
        Then,
        \begin{equation}
            \label{eq:av.forw.int}
            \bE \left[ \int_0^T \varphi(t) \, d^{-}B(t) \right]
            = \bE \left[ \int_0^T \bE \left[ D_{t^+} \, \varphi(t) | \cF_t \right] dt \right].
        \end{equation}
    \end{proposition}
    
    We also define the Donsker delta function~\cite{noep, do1} that we are frequently going to use in subsequent sections.
    \begin{definition}
        \label{ddfdefinition}
        Let $G: \Omega \to \bR$ be a random variable, with $G \in (\cS)^*$.
        The continuous function
        \begin{equation*}
            \delta_G \left( \cdot \right) : \bR \to \left(\cS\right)^*,
        \end{equation*}
        is called a \textit{Donsker delta function} of $G$ if it has the property that
        \begin{equation}
            \label{eq:delta1}
            \int_{\bR} f(g) \, \delta_G\left(g\right) \, dg = f\left(G\right),
        \end{equation}
        almost surely, for all measurable $f: \bR \to \bR$, such that the integral (understood in the sense of Bochner) converges.

        We call $\sD \subset L^2 \left( \Omega, \cF, P \right)$ the class of random variables for which the Donsker delta function exists.
    \end{definition}

We will also need the following result in the upcoming sections.

\begin{lemma}\label{lem:ddcond}
    Let $G\in\sD$ and let $Y$ be a $\cF_t$-measurable random variable with $\cF_t \subseteq \cF$. Let also $f: \bR^2 \to \bR$ be a measurable function which is, moreover, real analytic with infinite radius of convergence in its first argument. Then, the following equality
    \begin{equation}
        \label{eq:cond.av.donsker.delta}
        \bE[f(Y, G) | \cF_t]
        = \int_{\bR} f(Y, g) \, \bE[\delta_G(g)| \cF_t] \, dg 
    \end{equation}
    holds true provided $f(Y,G)$ is summable, where $\delta_G$ is the Donsker delta function of $G$.
\end{lemma}

\begin{proof}
    Any such function $f(y,g)$ can be written in terms of its Taylor expansion  $f(y,g)=\sum_{i,j} c_{ij} h_i(y) f_j(g)$. We have that, for any bounded $H\in\cF_t$,
    \begin{align*}
        \bE[H f(Y, G)]
        &= \bE[H  \sum_{i,j} c_{ij} h_i(Y) f_j(G)]
         = \sum_{i,j} c_{ij} \bE[H h_i(Y) f_j(G)] \\
        &= \sum_{i,j} c_{ij} \bE[H h_i(Y) \bE[f_j(G)| \cF_t]]
         = \bE[H \sum_{i,j} c_{ij} h_i(Y) \bE[f_j(G)| \cF_t]]
    \end{align*}
    where in the equation before the last, we used the fact that $H h_i(Y)\in\cF_t$.

    It follows that
    \begin{align*}
        \bE[f(Y, G) | \cF_t]
        &=\sum_{i,j} c_{ij} h_i(Y) \bE[f_j(G)| \cF_t]
        =\sum_{i,j} c_{ij} h_i(Y) \bE[\int_{\bR} f_j(g) \, \delta_G\left(g\right) \, dg| \cF_t] \\
        &= \int_{\bR} \sum_{i,j} c_{ij} h_i(Y) f_j(g) \, \bE[\delta_G\left(g\right) | \cF_t] \, dg
        =\int_{\bR} f(Y,g) \, \bE[\delta_G\left(g\right) | \cF_t] \, dg .
    \end{align*}
\end{proof}

    Since most of the analysis done in the next sections is based on the Wick product applied to smoothed white noise random variables, we collect in the next subsection some results related to its calculus.
    Some of them are original results, in particular Lemmata~\ref{cond.Donsker.delta} and~\ref{wprod.cond.expect}.

    \subsection{Wick products for smoothed white noise}
    Let $\omega_\phi$ be the \emph{smoothed white noise} as in~\eqref{smooth.wn} with variance  $||\phi||^2 = \int_0^T \phi^2_t dt$. We define the \emph{normalized} smoothed white noise as $\tilde\omega_\phi = \omega_\phi/||\phi||$.
    With $\psi\in L^1([0,T])$ another deterministic function, we define the \emph{non-centered} smoothed white noise as
    $\omega_{\phi,\psi} = \Psi + \omega_\phi$,
    with $\Psi=\int_0^T \psi_t dt$.
    We also define the normalized version $\tilde\omega_{\phi,\psi} = \omega_{\phi,\psi}/||\phi||$.
    Notice that $||\phi||^2$ is again the variance of $\omega_{\phi,\psi}$.

    The following result, originally due to It\^o~\cite{ito3}, expresses the Wick powers of Wiener integrals in terms of Hermite polynomials.

    \begin{lemma}
        \label{wick.power.weiner.int}
        Let $\omega_\phi$ be a smoothed white noise, then
        \begin{equation}
            \label{eq:wick.prod.omega.phi}
            (\omega_\phi)^{\wprod n}
            = ||\phi||^n h_n\left(\tilde\omega_\phi\right)
            =  ||\phi||^n \, \bE\left[\left(\tilde\omega_\phi + i Z\right)^n \Big| \cF_T\right].
        \end{equation}
    \end{lemma}
    \begin{proof}
        The first equality is taken from~\cite{oksendal2}, but see also equation (5.59) in~\cite{noep}, while the second follows since $h_n(x) = \bE[(x + i Z)^n]$.
    \end{proof}

    A similar result holds for the non-centered smoothed white noise.
    \begin{lemma}
        \label{wick.power.weiner.int.ext}
        Let $\omega_{\phi,\psi}$ be a non-centered smoothed white noise, then
        \begin{equation}
            \label{eq:wick.prod.omega.phi.ext}
            (\omega_{\phi,\psi})^{\wprod n}
            = ||\phi||^n h_n\left(\tilde\omega_{\phi,\psi}\right)
            =  ||\phi||^n \, \bE\left[\left(\tilde\omega_{\phi,\psi} + i Z\right)^n \Big| \cF_T\right].
        \end{equation}
    \end{lemma}
    \begin{proof}
        Since $h_n(x) = \bE[(x + i Z)^n]$, it is enough to prove the second equality.
        \begin{align*}
        (\omega_{\phi,\psi})
            ^{\wprod n}
            = \,  & (\Psi + \omega_\phi)^{\wprod n}
            = \sum_{k=0}^n \binom{n}{k} \Psi^k \omega_\phi^{\wprod n-k}                                                                                               \\
            = \,  & \sum_{k=0}^n \binom{n}{k} \Psi^k ||\phi||^{n-k} \, \bE\left[\left(\tilde\omega_\phi + i Z\right)^{n-k} \Big| \cF_T\right]                            \\
            = \,  & ||\phi||^n \bE\left[\sum_{k=0}^n \binom{n}{k} \left(\frac{\Psi}{||\phi||}\right)^k  \, \left(\tilde\omega_\phi + i Z\right)^{n-k} \Big| \cF_T\right] \\
            = \,  & ||\phi||^n \bE\left[\left(\frac{\Psi}{||\phi||} + \tilde\omega_\phi + i Z\right)^n \Big| \cF_T\right]
            = ||\phi||^n \bE\left[\left(\tilde\omega_{\phi,\psi} + i Z\right)^n \Big| \cF_T\right].
        \end{align*}
    \end{proof}

    By linearity, we can extend the previous result to the entire functions.
    \begin{lemma}
        Given the entire function
        $f(x)=\sum_{n\ge0} a_n x^n$, then
        \begin{equation}
            \label{eq:wick.entire.f}
            f^{\wprod}(\omega_{\phi,\psi})
            = \bE\left[f \left(\omega_{\phi,\psi} + i Z ||\phi||\right) \Big| \cF_T\right]
            = \bE\left[f \left((\tilde\omega_{\phi,\psi} + i Z)||\phi||\right) \Big| \cF_T\right],
        \end{equation}
        where $f^{\wprod}(\cX)=\sum_{n\ge0} a_n (\cX)^{\wprod n}$, $\cX \in \left(\cS\right)^*$
        and $Z$ is a standard Normal random variable independent of $\cF_T$.
    \end{lemma}
    \begin{proof}
        We prove the first equality.
        \begin{align*}
            f^{\wprod}(\omega_{\phi,\psi})
            =\sum_{n\ge0} a_n (\omega_{\phi,\psi})^{\wprod n}
            = \,  &\sum_{n\ge0} a_n ||\phi||^n \, \bE\left[\left(\tilde\omega_{\phi,\psi} + i Z\right)^n \Big| \cF_T\right] \\
            = \,  & \bE\left[\sum_{n\ge0} a_n  \left(\omega_{\phi,\psi} + i Z ||\phi||\right)^n \Big| \cF_T\right]          \\
            = \,  & \bE\left[f \left(\omega_{\phi,\psi} + i Z ||\phi||\right) \Big| \cF_T\right].
        \end{align*}
    \end{proof}

    \iffalse
    We give the following technical result that is not difficult to prove.
    \begin{proposition}
        Let $a>-1/2$, $b,c,d\in\bR$, $f(x)=\exp\left(a y^2 + b y +c\right)$ and $Z\sim\Normal(0,1)$, then with $i=\sqrt{-1}$,
        \begin{align}
            \bE\left[f\left(y + i d Z \right) \right]
            %= \,  & \frac{1}{\sqrt{1+2ad^2}}\exp\left(-\frac{(y/d-bd)^2}{2(1+2ad^2)}+\frac{(y/d)^2+2c}{2}\right) \\
            = \,  & \frac{1}{\sqrt{1+2ad^2}}\exp\left(-\frac{(y-bd^2)^2}{2d^2(1+2ad^2)}+\frac{y^2+2cd^2}{2d^2}\right).
        \end{align}
    \end{proposition}
    \fi

    Now, we present two useful results.
    \begin{lemma}
        \begin{equation}
            \label{eq:wprop.to.prod.power}
            \omega_{\phi,\psi}^{\wprod (n+1)} =
            \omega_{\phi,\psi} \cdot \omega_{\phi,\psi}^{\wprod n}  - n ||\phi||^2 \omega_{\phi,\psi}^{\wprod (n-1)}.
        \end{equation}
    \end{lemma}
    \begin{proof}
        By~\eqref{eq:wick.prod.omega.phi.ext},
        \begin{align*}
            \omega_{\phi,\psi}^{\wprod (n+1)}
            = \,  & ||\phi||^{n+1} h_{n+1}\left(\tilde\omega_{\phi,\psi}\right)                                                                                      \\
            = \,  & ||\phi||^{n+1} \tilde\omega_{\phi,\psi} h_n\left(\tilde\omega_{\phi,\psi}\right) - n ||\phi||^{n+1} h_{n-1}\left(\tilde\omega_{\phi,\psi}\right) \\
            = \,  & \omega_{\phi,\psi} \cdot \omega_{\phi,\psi}^{\wprod n}  - n ||\phi||^2 \omega_{\phi,\psi}^{\wprod (n-1)}  ,
        \end{align*}
        where in the second equality, we have applied the Hermite polynomial recurrence equation
        $h_{n+1}(x) = x \, h_n(x) - n h_{n-1}(x)$.
    \end{proof}

    The following result that is exploited in Lemma~\ref{cond.Donsker.delta}, has already appeared in Corollary~7.8 and Lemma~7.9 of~\cite{noep}, however here we give a different and more succinct proof.
    \begin{lemma}
        \label{lemma.wprop.to.prod.exp}
        Assume that $||\phi||^2\not=1$, then
        \begin{equation}
            \label{eq:wprop.to.prod.exp}
            \omega_{\phi,\psi} \wprod \exp^{\wprod}\left(-\frac{1}{2}(\omega_{\phi,\psi})^{\wprod2}\right)
            = \frac{1}{1-||\phi||^2} \omega_{\phi,\psi} \cdot \exp^{\wprod}\left(-\frac{1}{2}(\omega_{\phi,\psi})^{\wprod2}\right).
        \end{equation}
        If $||\phi||^2=1$, then
        $\omega_{\phi,\psi} \cdot \exp^{\wprod}\left(-\frac{1}{2}(\omega_{\phi,\psi})^{\wprod2}\right) = 0$.
    \end{lemma}
    \begin{proof}
        We have
        \begin{align*}
            & \omega_{\phi,\psi} \wprod \exp^{\wprod}\left(-\frac{1}{2}(\omega_{\phi,\psi})^{\wprod2}\right) \\
            & \quad = \omega_{\phi,\psi} \wprod \sum_{n\ge0} \frac{(-1)^n}{n!} \left(\frac{\omega_{\phi,\psi}}{\sqrt{2}}\right)^{\wprod2n}
            =\sqrt{2} \sum_{n\ge0} \frac{(-1)^n}{n!} \left(\frac{\omega_{\phi,\psi}}{\sqrt{2}}\right)^{\wprod2n+1}                                       \\
            & \quad = \sqrt{2}  \sum_{n\ge0} \frac{(-1)^n}{n!} \left(\frac{\omega_{\phi,\psi}}{\sqrt{2}} \cdot
            \left(\frac{\omega_{\phi,\psi}}{\sqrt{2}}\right)^{\wprod 2n}
            - n ||\phi||^2 \left(\frac{\omega_{\phi,\psi}}{\sqrt{2}}\right)^{\wprod (2n-1)}\right)                                                                 \\
            & \quad = \omega_{\phi,\psi} \cdot  \sum_{n\ge0} \frac{(-1)^n}{n!}
            \left(\frac{\omega_{\phi,\psi}}{\sqrt{2}}\right)^{\wprod 2n}
            + ||\phi||^2 \, \omega_{\phi,\psi}\wprod \sum_{n\ge1} \frac{(-1)^{n-1}}{(n-1)!} \left(\frac{\omega_{\phi,\psi}}{\sqrt{2}}\right)^{\wprod 2(n-1)} \\
            & \quad = \omega_{\phi,\psi} \cdot \exp^{\wprod}\left(-\frac{1}{2}(\omega_{\phi,\psi})^{\wprod2}\right)
            + ||\phi||^2 \, \omega_{\phi,\psi} \wprod \exp^{\wprod}\left(-\frac{1}{2}(\omega_{\phi,\psi})^{\wprod2}\right)  ,
        \end{align*}
        where in third equality we used~\eqref{eq:wprop.to.prod.power}.
        The result then follows.
    \end{proof}

    The random variable $B(T)$ admits the Donsker delta function, and we compute its conditional expectation with respect to $\cF_{(t-d)^+}$ in the following lemma.
    Since the conditional expectation of $B(T)$ with respect to this filtration is a Gaussian random variable, the result also directly follow by Proposition~7.2 of~\cite{noep}. % Eqn.~(7.16) of~\cite{noep}
    \begin{lemma}
        \label{cond.Donsker.delta}
        The conditional expectation of the Donsker delta function of $B(T)$ is given by the following formula
        \begin{equation}
            \bE \left[\delta_{B(T)}(g)| \cF_{(t-d)^+} \right]
            = \frac{1}{\sqrt{2 \pi T}} \exp^\wprod\left(-\frac{\left(g-B((t-d)^+)\right)^{\wprod 2}}{2T} \right).\label{eq:cond.donsker.delta}
        \end{equation}
    \end{lemma}
    \begin{proof}
        We have
        \begin{align*}
            \bE \left[(g-B(T))^{\wprod 2}| \cF_{(t-d)^+} \right]
            = \,  & \left(\bE \left[g-B(T)| \cF_{(t-d)^+} \right]\right)^{\wprod 2}
            = \left(g-B((t-d)^+)\right)^{\wprod 2}.
        \end{align*}
        Then, we get
        \begin{align*}
            \bE \left[\delta_{B(T)}(g)| \cF_{(t-d)^+} \right]
            = \,  & \bE \left[\frac{1}{\sqrt{2 \pi T}} \exp^{\wprod}
            \left(-\frac{(g-B(T))^{\wprod 2}}{2T}\right)| \cF_{(t-d)^+} \right]                                                      \\
            = \,  & \frac{1}{\sqrt{2 \pi T}} \exp^\wprod\left(-\frac{1}{2T} \bE \left[(g-B(T))^{\wprod 2}| \cF_{(t-d)^+} \right]\right) \\
            = \,  & \frac{1}{\sqrt{2 \pi T}} \exp^\wprod\left(-\frac{\left(g-B((t-d)^+)\right)^{\wprod 2}}{2T} \right).
        \end{align*}
    \end{proof}

    We finally compute a relation that allows to convert a Wick product in a normal product; it will be later used in Corollary~\ref{co:bs}.
    \begin{lemma}
        \label{wprod.cond.expect}
        \begin{align}
            \label{eq:wprod.cond.expect}
            & \frac{g-B((t-d)^+)}{T} \wprod \bE \left[\delta_{B(T)}(g)| \cF_{(t-d)^+} \right]
            = %\\ & \hspace{5cm} \notag
            \frac{g-B((t-d)^+)}{T-(t-d)^+} \, \bE \left[\delta_{B(T)}(g)| \cF_{(t-d)^+} \right].
        \end{align}
    \end{lemma}
    \begin{proof}
        By Lemma~\ref{cond.Donsker.delta}, we may write more explicitly the conditional expectation of the Donsker delta function, that is
        \begin{align*}
            & \kern-0.25cm
            \frac{g-B((t-d)^+)}{T} \wprod \bE \left[\delta_{B(T)}(g)| \cF_{(t-d)^+} \right]                                                                                                  \\
            = \,  & \frac{1}{\sqrt{T}}\frac{1}{\sqrt{2 \pi T}} \frac{g-B((t-d)^+)}{\sqrt{T}} \wprod \exp^\wprod\left(-\frac{1}{2}\left(\frac{g-B((t-d)^+)}{\sqrt{T}}\right)^{\wprod 2} \right).
        \end{align*}
        Then, by applying~\eqref{eq:wprop.to.prod.exp}, we get
        \begin{align*}
            & \kern-0.25cm \frac{g-B((t-d)^+)}{T} \wprod \bE \left[\delta_{B(T)}(g)| \cF_{(t-d)^+} \right]            \\
            = \,  & \frac{1}{\sqrt{2 \pi T^{3/2}}} \frac{g-B((t-d)^+)}{1-\Var\left[\frac{g-B((t-d)^+)}{\sqrt{T}}\right]} \cdot
            \exp^\wprod\left(-\frac{1}{2}\left(\frac{g-B((t-d)^+)}{\sqrt{T}}\right)^{\wprod 2} \right)                 \\
            %= \,  & \frac{1}{\sqrt{T}} \frac{1}{1-\frac{(t-d)^+}{T}}
            %\frac{g-B((t-d)^+)}{\sqrt{T}} \cdot \bE \left[\delta_{B(T)}(g)| \cF_{(t-d)^+} \right] \\
            = \,  & \frac{g-B((t-d)^+)}{T-(t-d)^+} \, \bE \left[\delta_{B(T)}(g)| \cF_{(t-d)^+} \right].
        \end{align*}
    \end{proof}

    \iffalse
    We define the following function
    \begin{equation}
        \label{eq:phi.wprod}
        \phi^{\wprod}_{\omega_{\phi,\psi}}
        = \frac{1}{\sqrt{2\pi}||\phi||}
        \exp^{\wprod}\left(-(\tilde\omega_{\phi,\psi})^{\wprod2}/2\right).
    \end{equation}
    \fi

    \section{The Black-Scholes model}\label{sec:bsm}

    The Black-Scholes model~\cite{bsm, hull, lambertonlapeyre} describes the behavior of the market prices with a continuous-time portfolio composed by one riskless and one risky asset. As mentioned in the Introduction, it assumes the behavior of the riskless asset $X_0(t)$, aka the bond, to be given by the ordinary differential equation~\eqref{eq:cbsm1a}-\eqref{eq:cbsm1b}, and the dynamics of the risky asset $X_1(t)$, aka the stock, to be described by the stochastic differential equation~\eqref{eq:cbsm2a}-\eqref{eq:cbsm2b}. We pose this problem in the probability space specified in the Introduction.
    
    We assume the presence of an AIT, who knows at time $t\in[0,t]$ the value of $G\in\sD$ together with the information $\cF_{(t-d)^+}$ where $d\in (0,T]$ is a constant that represents the delay in the information flow of the stock.
    $G$ is a random variable assumed to be measurable with respect to the $\cF_T$, that is, the sigma-algebra generated by all values of $(B(t), t\in[0,T])$.
    Thus, the trader can choose portfolios from the class of $\bG$-predictable processes, where $\bG = \lbrace \cG_t \rbrace _{t\in [0,T]}$ and
    \begin{equation*}
        \cG_t
        = \lbrace \sigma(B(s)): 0 \leq s \leq \left(t-d\right)^{+} \rbrace \vee \sigma \left(G \right)
        = \cF_{(t-d)^+} \vee \sigma \left(G \right).
    \end{equation*}
    Let us consider the wealth process $X^{\pi}(t)$ satisfying the stochastic differential equation
    \begin{align}
        \label{eq:bsm}
        dX^{\pi}(t) = \,  &
        \left( 1 - \pi \left(t, G\right) \right) \rho(t) \, X^{\pi}(t) \, dt \\ &
        + \pi \left(t, G\right) X^{\pi}(t) \left( \mu(t) \, dt + \sigma(t) \, d^{-}B(t) \right), \nonumber
    \end{align}
    where we have assumed, w.l.o.g., $X^{\pi}(0) = 1$,
    and where $\rho(t)$, $\mu(t)$, and $\sigma(t)$ are deterministic continuous bounded functions, defined as before, with $\sigma(t)$ being bounded away from zero.
    \begin{definition}
        \label{def:bs1}
        Let $\cA$ be the set of self-financing portfolios $\pi \left(t, G\right)$ such that
        \begin{enumerate}[(i)]
            \item the control processes $\pi \in \cA$ are c\`agl\`ad and $\bG$-adapted such that for all $t\in [0,T]$ and $g \in \bR$, $\pi (t,g)$ is $\cF_{(t-d)^+}$-measurable,
            \item the product $\sigma \pi$ is c\`agl\`ad and forward integrable with respect to $B(t)$,
            \item for all $\pi \in \cA$,
            $\bE \left[ \int_0^T \pi^2(t,G) \, dt \right] < \infty$.
        \end{enumerate}
    \end{definition}
    Once defined the above, the aim is to find the optimal portfolio $\hat{\pi} \in \cA$ that maximizes the expected logarithm of the final wealth, such that
    \begin{align*}
        \cV^{\hat{\pi}} & \coloneqq \bE \left[ \log \left(X^{\hat{\pi}}(T) \right) \right]
        = \sup_{\pi \in \cA} \bE \left[ \log \left(X^{\pi}(T) \right) \right],
    \end{align*}
    and the market is said to be viable if $\cV^{\hat{\pi}} < \infty$.
    \begin{theorem}
        \label{th:bs1}
        Let us consider the financial market given by model~\eqref{eq:bsm}, where the AIT has, for all $t \in [0,T]$, access to the information $\cG_t$.
        With $g\in\supp(G)$, let
        \begin{equation}
            \label{eq:alpha}
            \alpha_d(t,g) \coloneqq \frac{\bE \left[ D_{t^+} \, \delta_G(g)| \cF_{(t-d)^+} \right]}{\bE \left[ \delta_G(g)| \cF_{(t-d)^+} \right]},
        \end{equation}
        then the optimal strategy divergence, $\Delta\hat\pi(t,G)
        \coloneqq \hat\pi(t,G) - \bar\pi(t)$,  is equal to
        \begin{equation}
            \label{eq:pi.bsm}
            \Delta\hat\pi(t,g)  = \alpha_d(t,g)/\sigma(t).
        \end{equation}
        In the strategy divergence definition, $\hat\pi(t,G)$ denotes the AIT optimal portfolio and $\bar{\pi}(t)$ is the classical Black-Scholes-Merton strategy given in~\eqref{eq:merton.pi}.

            Furthermore, the additional value of the expected logarithm of the final wealth,
        $\Delta \cV^{\hat{\pi}} \coloneqq \cV^{\hat{\pi}} - \cV^{\bar{\pi}}$  is equal to
        \begin{align}
            \Delta \cV^{\hat{\pi}} \nonumber
            = \, & \bE \Bigg[  \int_0^T  \, \alpha_d(t,g) \, d^{-}B(t) -
            \int_0^T \frac{1}{2} \alpha^2_d(t,G) \, dt  \Bigg]                                                \\
            = \, & \int_0^T \bE \left[ D_{t^+} \left( \alpha_d(t,g) \right) - \frac{1}{2} \alpha^2_d(t,G) \right] \, dt, \label{E.logX.BSM}
        \end{align}
        with $\cV^{\bar{\pi}} \coloneqq \bE \left[ \log \left(X^{\bar{\pi}}(T) \right) \right] = \int_0^T \left(\frac{1}{2} \sigma^2(t) \bar\pi^2(t) + \rho(t) \right) dt$ being the expected logarithm of the final wealth of the classical Black-Scholes-Merton portfolio.
    \end{theorem}
    \begin{proof}
        The Russo-Vallois forward integral preserves It\^o calculus under suitable assumptions, see Theorem~8.12 in~\cite{noep}.
        Hence, if $\pi \in \cA$, the solution $X^{\pi}(t)$, $t \in [0,T]$, of~\eqref{eq:bsm}, assuming $X(0)=1$, is given by
        \begin{align*}
            X^{\pi}(t)
            = \,  & \exp \bigg \lbrace \int_0^t \left( \left( \mu(s) - \rho(s) \right) \pi (s, G) + \rho(s) - \frac{1}{2} \, \sigma^2(s) \, \pi^2(s, G) \right) ds \\
            & + \int_0^t \sigma(s) \, \pi (s, G) \, d^{-}B(s) \bigg \rbrace.
        \end{align*}
        In order to find the optimal portfolio $\hat{\pi} \in \cA$ that maximizes the expected logarithm of the final wealth, we compute $\cV^\pi= \bE \left[ \log \left(X^{\pi}(T) \right) \right]$ explicitly to find
        \begin{align}
            \cV^\pi %\bE \left[ \log \left(X^{\pi}(T) \right) \right]
            = \,  & \bE \bigg[ \int_0^T \left( \left( \mu(t) - \rho(t) \right) \pi (t, G) + \rho(t) - \frac{1}{2} \, \sigma^2(t) \, \pi^2(t, G) \right) dt \label{eq:bsm2} \\
            & \quad  + \int_0^T \sigma(t) \, \pi (t, G) \, d^{-}B(t) \bigg]. \nonumber
        \end{align}
        Using the tower property and the expectation of the forward integral, Equation~\eqref{eq:av.forw.int}, we get
        \begin{align*}
            \cV^\pi %\bE \left[ \log \left(X^{\pi}(T) \right) \right]
            = \,  & \bE \bigg[ \int_0^T \Big( \left( \mu(t) - \rho(t) \right) \bE \left[ \pi (t, G)|\cF_{(t-d)^+} \right] + \rho(t) \\
            & \quad - \frac{1}{2} \, \sigma^2(t) \, \bE \left[ \pi^2 (t, G)|\cF_{(t-d)^+} \right]                               \\
            & \quad + \sigma(t) \, \bE \left[ D_{t^+} \, \pi(t, G) | \cF_{(t-d)^+} \right] \Big) \, dt \, \bigg].
        \end{align*}
        At this point, we use the definition of the Donsker delta function, see Definition~\ref{ddfdefinition}, and apply~\eqref{eq:cond.av.donsker.delta} from Lemma~\ref{lem:ddcond} to get
        \begin{align*}
            \cV^\pi
            = \,  & \bE \bigg[ \int_0^T\int_{\bR} \Big( \big( \left( \mu(t) - \rho(t) \right)\pi (t,G) + \rho(t)                       \\
            & \quad - \frac{1}{2} \, \sigma^2(t) \, \pi^2 (t, G) \big) \, \bE \left[ \delta_G(G)|\cF_{(t-d)^+} \right]             \\
            & \quad + \sigma(t) \, \pi(t, G) \, \bE \left[ D_{t^+} \, \delta_G(G)| \cF_{(t-d)^+} \right] \Big) \, dg \, dt \bigg].
        \end{align*}
        Then, maximizing the integrand for all $t \in [0,T]$ and $g \in \bR$, the optimal portfolio $\hat{\pi}(t,g)$ is given by~\eqref{eq:pi.bsm}.
        Substituting $\hat{\pi}(t,g) = \bar\pi(t) +  \alpha_d(t,g)/\sigma(t)$ into Equation~\eqref{eq:bsm2}, we get
        \begin{align*}
            \cV^\pi
            = \, &\bE \Bigg[ \int_0^T \left(\frac{1}{2} \sigma^2(t) \bar\pi^2(t)+ \rho(t) \right) dt
            - \int_0^T \frac{1}{2} \alpha^2_d(t,G) dt
            + \int_0^T  \, \alpha_d(t,G) \, d^{-}B(t) \Bigg]                                       \\
            = \, &\bE \Bigg[ \int_0^T \left(\frac{1}{2} \sigma^2(t) \bar\pi^2(t)+ \rho(t) \right) dt
            - \int_0^T \frac{1}{2} \alpha^2_d(t,G) dt
            + \int_0^T D_{t^+} \left( \alpha_d(t,G) \right)  dt  \Bigg].
        \end{align*}
        In the first equation we used the fact that $\int_0^t \sigma(t) \bar\pi(t) \, d^{-}B(t)$ is a martingale with null expectation, as the classical strategy $\bar\pi(t)$ is deterministic (adapted would be enough).
        Equation~\eqref{E.logX.BSM} follows after interpreting the expectation of the first integral as the classical portfolio value in the Black-Scholes-Merton model.
    \end{proof}

    \begin{corollary}
        \label{co:bs}
        If the insider information is $G=B(T)$, then the strategy divergence is
        \begin{equation}
            \Delta\hat\pi(t,B(T)) %\coloneqq \hat\pi(t,B(T)) - \bar\pi(t)
            = \frac{1}{\sigma(t)} \frac{B(T) - B(\left(t-d\right)^{+})}{T - \left( t - d \right)^{+}}
            , \label{eq:delta.pi.bsm}
        \end{equation}
        the market is viable for any positive $d$,
        and the additional value of the expected logarithm of the final wealth is, for $d\in(0,T]$,
        \begin{align}
            \Delta \cV^{\hat{\pi}}(T)
            = \frac{d}{2 \, T} + \frac{1}{2} \ln \left( \frac{T}{d} \right) > 0. \label{delta.log.BSM}
        \end{align}
        Then, the AIT obtains more expected utility than the traditional trader despite the presence of the delay.
    \end{corollary}

    \begin{proof}
        We have $\Delta\hat\pi(t,g)=\alpha_d(t,g)/\sigma(t)$ with
        \begin{equation*}
            \alpha_d(t,g) \coloneqq \frac{\bE \left[ D_{t^+} \, \delta_{B(T)}(g)| \cF_{(t-d)^+} \right]}{\bE \left[ \delta_{B(T)}(g)| \cF_{(t-d)^+} \right]}.
        \end{equation*}

        We have that $\delta_{B(T)}(g)\in\left(\cS\right)^*$ for any $g\in\bR$,  and it is defined as
        \begin{equation*}
            \delta_{B(T)}(g)
            %= \frac{1}{\sqrt{2 \pi T}} \exp^{\wprod} \left(-\frac{(g-B(T))^{\wprod 2}}{2T}\right)
            = \frac{1}{\sqrt{2 \pi T}} \exp^{\wprod} \left(-\frac{Y_g^{\wprod 2}}{2}\right)
            %=: \phi^\wprod_{Y_g},
        \end{equation*}
        where $Y_g=(g-B(T))/\sqrt{T}$.

        By the Wick chain rule, see Proposition 5.14 in~\cite{noep}, we have
        \begin{equation*}
            D_{t^+} \, \delta_{B(T)}(g)
            = \delta_{B(T)}(g) \wprod D_{t^+} \!\! \left(-\frac{1}{2}Y_g^{\wprod 2}\right) \!
            = -\delta_{B(T)}(g) \wprod Y_g \wprod D_{t^+} Y_g
            = \frac{1}{\sqrt{T}} \delta_{B(T)}(g) \wprod Y_g ,
        \end{equation*}
        where in the last equation we used the fact that $D_{t^+} Y_g=-1/\sqrt{T}$.

        By the distributive property of the conditional expectation with respect to the Wick product, see Lemma 6.20 in~\cite{noep}, we have
        \begin{align*}
            \bE \left[ D_{t^+} \, \delta_{B(T)}(g)| \cF_{(t-d)^+} \right]
            = \,  & \frac{1}{\sqrt{T}} \bE \left[\delta_{B(T)}(g)| \cF_{(t-d)^+} \right]  \wprod \bE \left[Y_g | \cF_{(t-d)^+} \right] \\
            = \,  & \bE \left[\delta_{B(T)}(g)| \cF_{(t-d)^+} \right]  \wprod  \frac{g-B((t-d)^+)}{T}                                  \\
            = \,  & \frac{g-B((t-d)^+)}{T-(t-d)^+} \, \bE \left[\delta_{B(T)}(g)| \cF_{(t-d)^+} \right],
        \end{align*}
        where in the last equality we used the results of Lemma~\ref{wprod.cond.expect}.
        From that immediately follows that
        \begin{align*}
            \alpha_d(t,g) = \frac{g-B((t-d)^+)}{T-(t-d)^+} ,
        \end{align*}
        and~\eqref{eq:delta.pi.bsm} holds.

        For the additional value of the expected logarithm of the final wealth, from~\eqref{E.logX.BSM}, we have
        $
            \Delta \cV^{\hat{\pi}}
            = \bE \left[  \int_0^T D_{t^+} \left( \alpha(t,B(T)) \right)  - \frac{1}{2} \alpha^2_d(t,B(T)) dt  \right].
        $
        Since, for $t\in[0,T]$,
        \begin{equation*}
            D_{t^+} \left( \alpha_d(t,B(T)) \right)
            = D_{t^+} \, \frac{B(T) - B(\left(t-d\right)^{+})}{T - \left( t - d \right)^{+}}
            = \frac{1}{T - \left( t - d \right)^{+}} ,
        \end{equation*}
        we finally get that
        \begin{align*}
            \Delta \cV^{\hat{\pi}}
            = \,  & \bE \left[  \int_0^T \frac{1}{T - \left( t - d \right)^{+}}  - \frac{1}{2}\left(\frac{B(T) - B(\left(t-d\right)^{+})}{T - \left( t - d \right)^{+}} \right)^2 dt \right] \\
            = \,  &  \frac{d}{2 \, T} + \frac{1}{2} \ln \left( \frac{T}{d} \right),
        \end{align*}
    which is, obviously, finite and positive for the parameter values under consideration.
    \end{proof}

    \section{The Heston model}\label{sec:hes}

    The Heston model was introduced in Section~\ref{sec:fm}; as mentioned there, it assumes a stochastic volatility given by the solution to the stochastic differential equation~\eqref{eq:hes1a}-\eqref{eq:hes1b}. In this section, we employ this model and further assume that the AIT knows at time $t\in[0,T]$ the value $G\in\sD$, measurable in $\cF_T$, in addition to the filtration $\cF_{\left(t-d\right)^{+}}$ for $d\in(0,T]$ constant, but possesses no future information about the stock volatility.
    The allowed portfolios, therefore, belong to the class of $\bG$-predictable processes, where $\bG = \lbrace \cG_t \rbrace _{t\in [0,T]}$ is given by
    \begin{align*}
        \cG_t = \,  & \lbrace \sigma(B(s)): 0 \leq s \leq \left(t-d\right)^{+} \rbrace \vee \lbrace \sigma(W(s)): 0 \leq s \leq t \rbrace \vee \sigma(G)
        \\
        = \,  & \cF_{(t-d)^+} \vee \cH_t \vee \sigma(G).
    \end{align*}
    Let us consider that the wealth process $X^{\pi}(t)$ satisfies the stochastic differential equation
    \begin{align}
        \label{eq:hes}
        dX^{\pi}(t)
        = \,  & \left( 1 - \pi \left(t, G\right) \right) \rho(t) \, X^{\pi}(t) \, dt                              \\
        & + \pi \left(t, G\right) X^{\pi}(t) \left( \mu(t) \, dt + \sqrt{V(t)} \, d^{-}B(t) \right), \nonumber
    \end{align}
    where we assume, w.l.o.g., $X^{\pi}(0) = 1$.
    The deterministic parameters $\rho(t)$ and $\mu(t)$ are as in the previous section, while the volatility process $V(t)$ satisfies the stochastic differential equation~\eqref{eq:hes1a}-\eqref{eq:hes1b}, and consequently it is positive for all $t\in [0,T]$ by Proposition~\ref{prop:cir}.
    The set of self-financing portfolios $\pi \in \cA$ are characterized according to Definition~\ref{def:bs1}. The main result of this section comes as follows.

    \begin{theorem}
        \label{th:hes1}
        Let us consider the financial market given by model~\eqref{eq:hes}, where the AIT has, for all $t \in [0,T]$, access to the information $\cG_t$.
        The optimal strategy divergence is
        \begin{equation}
            \label{eq:heston.pi}
            \Delta\hat{\pi}(t,g) = \alpha_d(t,g)/\sqrt{V(t)},
        \end{equation}
        where $\alpha_d(t,g)$ is defined as in~\eqref{eq:alpha} and the optimal portfolio of the traditional trader is $\bar\pi(t) = (\mu(t) - \rho(t))/V(t)$, $t \in [0,T]$.

        The additional value of the expected logarithm of the final wealth is again given by~\eqref{E.logX.BSM},
        and the value of the expected logarithm of the final wealth for the traditional trader is given by
        \begin{equation*}
            \cV^{\bar \pi} =  \int_0^T \left( \frac{1}{2} \left( \mu(t) - \rho(t) \right)^2\, \bE\left[\frac{1}{V(t)}\right] + \rho(t) \right) dt,
        \end{equation*}
        which is finite provided $\kappa \theta \geq \eta^2$.
    \end{theorem}
    \begin{proof}
        With $X^\pi(0)=1$ and $\pi \in \cA$, the solution $X^\pi(t)$, $t \in [0,T]$, of~\eqref{eq:hes}, is
        \begin{align*}
            X^{\pi}(t)
            = \,  & \exp \bigg \lbrace \int_0^t \left( \left( \mu(s) - \rho(s) \right) \pi (s, G) + \rho(s) - \frac{1}{2} \, V(s) \, \pi^2(s, G) \right) ds \\
            & + \int_0^t \sqrt{V(s)} \, \pi (s, G) \, d^{-}B(s) \bigg \rbrace.
        \end{align*}
        We compute the expected logarithm of the final wealth, $\cV^\pi = \bE \left[ \log \left(X^{\pi}(T)\right) \right]$, to find
        \begin{align}
            \label{eq:hes3}
            \cV^\pi
            = \,  & \bE \bigg [ \int_0^T \left( \left( \mu(t) - \rho(t) \right) \pi (t, G) + \rho(t) - \frac{1}{2} \, V(t) \, \pi^2(t, G) \right) dt \\
            & +  \int_0^T \sqrt{V(t)} \, \pi (t, G) \, d^{-}B(t) \bigg]. \nonumber
        \end{align}
        By conditioning on $\cF_{(t-d)^+} \vee \cH_t$, after applying the tower property and then using formulas~\eqref{eq:av.forw.int} and~\eqref{eq:cond.av.donsker.delta}, we have
        \begin{align*}
            \cV^\pi
             = \,  & \bE \bigg [  \int_0^T \int_\bR \Big( \left( \mu(t) - \rho(t) \right) \pi (t,g) \, \bE\left[\delta_G(g) | \cF_{(t-d)^+} \vee \cH_t \right] + \rho(t) \\
            &  - \frac{1}{2} \, V(t) \, \pi^2 (t,g) \, \bE \left[ \delta_G(g) | \cF_{(t-d)^+} \vee \cH_t \right]                                           \\
            & + \sqrt{V(t)} \, \pi (t, g) \, \bE \left[ D_{t^+} \, \delta_G(g)  | \cF_{(t-d)^+} \vee \cH_t \right] \Big) \, dg  \, dt \nonumber
            \bigg ] \\
            = \,  & \bE \bigg [  \int_0^T \int_\bR \Big( \left( \mu(t) - \rho(t) \right) \pi (t,g) \, \bE\left[\delta_G(g) | \cF_{(t-d)^+} \right] + \rho(t) \\
            & - \frac{1}{2} \, V(t) \, \pi^2 (t,g) \, \bE \left[ \delta_G(g) | \cF_{(t-d)^+} \right]                                           \\
            & + \sqrt{V(t)} \, \pi (t, g) \, \bE \left[ D_{t^+} \, \delta_G(g)  | \cF_{(t-d)^+} \right] \Big) \, dg   \, dt \nonumber
            \bigg ],
        \end{align*}
        where we used the fact that $G$ is independent of $W(t)$ and that $\pi^2 (t,g)$ is $\cF_{(t-d)^+} \vee \cH_t$ measurable.
        By optimizing for each $\omega\in\Omega$, for all $t \in [0,T]$ and $g \in \bR$, the optimal portfolio $\hat{\pi}(t,g)$ is then equal to
        \begin{equation*}
            \hat{\pi}(t,g) = \frac{\mu(t) - \rho(t)}{V(t)} + \frac{\alpha_d(t,g)}{\sqrt{V(t)}},
        \end{equation*}
        and~\eqref{eq:heston.pi} holds true.
        The rest follows as in the proof of Theorem~\ref{th:bs1}. Note, in particular, that the optimal portfolio of the traditional trader is viable as a consequence of Proposition~\ref{prop:cir} provided $\kappa \theta \geq \eta^2$, since this implies that $\bE\left[1/V(t)\right]<\infty$.
        Note also that this condition is stronger than the Feller condition, so $V(t)$ always stays positive and thus all the denominators are well-defined (see, again, the statement of Proposition~\ref{prop:cir}).
    \end{proof}

    The next corollary follows from the results of Theorem~\ref{th:hes1} and a similar proof of Corollary~\ref{co:bs}.

    \begin{corollary}
        \label{co:hes}
        Consider the financial market given by the model~\eqref{eq:hes} with $\kappa \theta \geq \eta^2$ and $G=B(T)$.
        The same conclusions can be drawn to those found in Corollary~\ref{co:bs} for the Black-Scholes-Merton model.
        The AIT gets, in average, more utility than the traditional trader, and the difference in the expected logarithm of the final wealth is given in~\eqref{delta.log.BSM}.
    \end{corollary}

    \section{The Vasicek model}\label{sec:vk}

    The Vasicek model assumes fluctuations in the interest rate as described in Section~\ref{sec:fm}. Herein, we adopt this model and also assume the AIT knows at time $t\in[0,T]$ the value $G\in\sD$, measurable in $\cF_T$, in addition to the filtration $\cF_{\left(t-d\right)^{+}}$ for $d\in(0,T]$ constant, and no future information about the interest rate, that is, only knows $\cH_t$ (see the Introduction for the definition of this filtration).

    In consequence, the portfolios belong to the class of $\bG$-predictable processes, where $\bG =\lbrace \cG_t \rbrace _{t\in [0,T]}$ with
    \begin{align*}
        \cG_t = \,  & \lbrace \sigma(B(s)): 0 \leq s \leq \left(t-d\right)^{+} \rbrace \vee \lbrace \sigma(W(s)): 0 \leq s \leq t \rbrace \vee \sigma(G) \\
        = \,  & \cF_{(t-d)^+} \vee \cH_t \vee \sigma(G).
    \end{align*}
    Let us consider the wealth process $X^{\pi}(t)$ satisfying the stochastic differential equation
    \begin{align}
        \label{eq:vk2}
        dX^{\pi}(t) = \,  & \left( 1 - \pi \left(t, G\right) \right) R(t) \, X^{\pi}(t) \, dt                                \\
        & + \pi \left(t, G\right) X^{\pi}(t) \left( \mu(t) \, dt + \sigma(t) \, d^{-}B(t) \right),  \nonumber
    \end{align}
    again with $X^{\pi}(0) =1$ and where $\mu(t)$ and $\sigma(t)$ are deterministic continuous bounded functions, defined as in the previous sections, with the volatility satisfying the condition $||1/\sigma(\cdot)||_{\infty}<\infty$ (that is, it is bounded away from zero).
    The short-rate process $R(t)$ obeys the stochastic differential equation~\eqref{eq:vk1a}-\eqref{eq:vk1b}.
    The set of self-financing portfolios $\pi \in \cA$ are the same as in Definition~\ref{def:bs1}.
    \begin{theorem}
        \label{th:vk1}
        Let us consider the financial market given by the model~\eqref{eq:vk2}, where the AIT has, for all $t \in [0,T]$, access to the information $\cG_t$.
        The optimal strategy divergence is the same as in the Black-Scholes-Merton model~\eqref{eq:pi.bsm} and the optimal portfolio of the traditional trader is
        $\bar\pi(t) = (\mu(t) - R(t))/\sigma^2(t)$, $t \in [0,T]$.

        Moreover, the additional value of the expected logarithm of the final wealth is given by~\eqref{E.logX.BSM},
        and the value of the expected logarithm of the final wealth for the traditional trader read
        \begin{equation}
            \cV^{\bar \pi}
            = \int_0^T \left( \frac{\bE[\left(\mu(t) - R(t)\right)^2]}{2 \, \sigma^2(t)} + \bE\left[R(t)\right] \right) dt. \label{eq:V.vk}
        \end{equation}
    \end{theorem}
    \begin{proof}
        With $X^{\pi}(0)=1$ and $\pi \in \cA$, the solution $X^{\pi}(t)$, $t \in [0,T]$, of~\eqref{eq:vk2}, is given by
        \begin{align}
            X^{\pi}(t)
            = \,  & \exp \bigg \lbrace \int_0^t \left( \left( \mu(s) - R(t) \right) \pi (s, G) + R(t) - \frac{1}{2} \, \sigma^2(s) \, \pi^2(s, G) \right) ds \nonumber \\
            & + \int_0^t \sigma(s) \, \pi (s, G) \, d^{-}B(s) \bigg \rbrace. \label{eq:vk4}
        \end{align}
        We compute the expected logarithm of the final wealth to find
        \begin{align}
            \bE \left[ \log \left(X^{\pi}(T)\right) \right]
            = \,  & \bE \left[ \int_0^T \left( \left( \mu(t) - R(t) \right) \pi (t, G) + R(t) - \frac{1}{2} \, \sigma(t)^2 \, \pi^2(t, G) \right) dt \right.
            \nonumber                                                                                                                                     \\
            & + \left. \int_0^T \sigma(t) \, \pi (t, G) \, d^{-}B(t) \right]. \label{eq:vk5}
        \end{align}
        By means the tower property and the expectation of the forward integral, we get
        \begin{align*}
            \bE \left[ \log \left(X^{\pi}(T) \right) \right]
            = \,  & \bE \Bigg[ \int_0^T \int_\bR \Big( \left(\mu(t) - R(t) \right) \pi (t, g) \, \bE \left[ \delta_G(g) | \cF_{(t-d)^+} \right] + R(t) \\
            & - \frac{1}{2} \, \sigma^2(t) \, \pi^2 (t, g) \, \bE \left[ \delta_G(g) | \cF_{(t-d)^+} \right]                              \\
            & + \sigma(t) \, \pi (t,g) \, \bE \left[ D_{t^+} \, \delta_G(g) | \cF_{(t-d)^+} \right] \Big) \, dg \, dt \Bigg].
        \end{align*}
        Following the same argument as in Theorem~\ref{th:bs1} we have that,
        for all $t \in [0,T]$ and $g \in \bR$, the optimal portfolio $\hat{\pi}(t,g)$ is
        \begin{equation*}
            \hat{\pi}(t,g) = \frac{\mu(t) - R(t)}{\sigma^2(t)} +  \frac{\alpha_d(t,g)}{\sigma(t)} ,
        \end{equation*}
        with $\alpha_d(t,g)$ defined in~\eqref{eq:alpha}.

        Substituting $\hat{\pi}(t,g)$ into equation~\eqref{eq:vk5}, the expected logarithm of the final wealth is
        \begin{equation*}
            \bE\left[\log \left(X^{\hat{\pi}}(T) \right) \right]
            = \int_0^T \left( \frac{\bE[\left(\mu(t) - R(t)\right)^2]}{2 \, \sigma^2(t)} + \bE\left[R(t)\right] \right) dt
            + \Delta \cV^{\hat{\pi}} ,
        \end{equation*}
        where $\Delta \cV^{\hat{\pi}}$ is defined in~\eqref{E.logX.BSM}.
        From Section~\ref{sec:fm}, we know that the expectation terms $\bE[R(t)]$ and $\bE[R(t)^2]$ are finite, and therefore the market for the traditional trader is viable.
    \end{proof}

    \begin{corollary}
        \label{co:vk}
        Let us consider the financial market given by model~\eqref{eq:vk2}.
        The same conclusions can be drawn to those found in Corollary~\ref{co:bs} for the Black-Scholes-Merton model.
        The AIT always gets more expected utility than the traditional trader and the difference in the expected logarithm of the final wealth is given by~\eqref{delta.log.BSM}.
    \end{corollary}

    \begin{remark}
        We have shown, under equivalent hypotheses, that the obtained results coincide for all the models considered if the delay~$d$ is present only in the information flow of the stock.
        However, if we consider that any of the other information flows is delayed too, the advantage of the AIT over the traditional trader is not necessarily guaranteed. Showing this is the goal of the next section.
    \end{remark}

    \section{The Vasicek model with delay in two information flows}\label{sec:vk.delay}
    
    Lastly, we consider the Vasicek model, as in the previous section, but now we assume that there exists a delay in both information flows, that is, the stock and the interest rate dynamics.
    Thus, the AIT can choose portfolios from the class of $\bG$-predictable processes, where $\bG =\lbrace \cG_t \rbrace _{t\in [0,T]}$ is given by
    \begin{align*}
        \cG_t = \,  & \lbrace \sigma(B(s)): 0 \leq s \leq \left(t-d\right)^{+} \rbrace \vee \lbrace \sigma(W(s)): 0 \leq s \leq \left(t-d\right)^{+} \rbrace \vee \sigma \left(G \right) \\
        = \,  & \cF_{(t-d)^+} \vee \cH_{(t-d)^+} \vee \sigma \left(G \right).
    \end{align*}
    The wealth process $X^{\pi}(t)$ obeys the stochastic differential equation
    \begin{align}
        \label{eq:vk9}
        dX^{\pi}(t)
        = \,  & \left( 1 - \pi \left(t, G\right) \right) R(t) \, X^{\pi}(t) \, dt                               \\
        & + \pi \left(t, G\right) X^{\pi}(t) \left( \mu(t) \, dt + \sigma(t) \, d^{-}B(t) \right), \nonumber
    \end{align}
    with $X^{\pi}(0)=1$ and where the parameters $\mu(t)$ and $\sigma(t)$, the stochastic process $R(t)$, and the class of admissible control processes $\pi \in \cA$ are as in Section~\ref{sec:vk}.

    \begin{theorem}
        \label{th:vk3}
        Let us consider the financial market given by model~\eqref{eq:vk9}, where the AIT has access to the information $\cG_t = \cF_{(t-d)^+} \vee \cH_{(t-d)^+} \vee \sigma \left(G \right)$, for all $t \in [0,T]$.
        The optimal portfolio $\hat{\pi} \in \cA$, with $g\in\supp(G)$, is
        \begin{equation*}
            \hat{\pi}(t,g)
            = \frac{\mu(t) - \bE\left[R(t)|\cH_{(t-d)^+}\right]}{\sigma^2(t)} + \frac{\alpha_d(t,g)}{\sigma(t)},
        \end{equation*}
        and the additional value of the expected logarithm of the final wealth is equal to
        \begin{equation*}
            \Delta\cV^{\hat{\pi}}
            = \int_0^T \bE\left[\frac{\Var \left[R(t)|\cH_{(t-d)^+}\right]}{2 \, \sigma^2(t)}
            + D_{t^+} \left( \alpha_d(t,G) \right) - \frac{1}{2} \alpha^2_d(t,G) \right] \, dt.
        \end{equation*}
    \end{theorem}
    \begin{proof}
        If $\pi \in \cA$, the solution $X^{\pi}(t)$, $t \in [0,T]$, of~\eqref{eq:vk2}, is given by equation~\eqref{eq:vk4}.
        We compute the expected logarithm of the final wealth explicitly to find
        \begin{align}
            \label{eq:vk10}
            \cV^{\pi}
            = \,  &  \bE \bigg[\int_0^T \left( \left( \mu(t) - R(t) \right) \pi (t, G) + R(t) - \frac{1}{2} \, \sigma^2(t) \, \pi^2(t, G) \right) dt  \\
            & + \int_0^T \sigma(t) \, \pi (t, G) \, d^{-}B(t) \bigg]  \nonumber\\
            = \,  & \bE\bigg[\int_0^T \Big(\big( \mu(t) - \bE \left[ R(t) |\cH_{(t-d)^+}\right] \big) \, \bE \left[ \pi (t, G) |\cF_{(t-d)^+}\right]   \nonumber \\
            & + \bE \left[ R(t) |\cH_{(t-d)^+}\right] - \frac{1}{2} \, \sigma^2(t) \, \bE \left[ \pi^2 (t, G) |\cF_{(t-d)^+}\right] \nonumber\\
            & + \sigma(t) \, \bE \left[ D_{t^+} \, \pi (t, G) \right] \Big) \, dt \bigg], \nonumber
        \end{align}
        where, in the second equality, we used the tower law for the expectation conditioned on~$\cF_{(t-d)} \vee \cH_{(t-d)^+}$ and we applied~\eqref{eq:av.forw.int}.
        By~\eqref{eq:cond.av.donsker.delta} from Lemma~\ref{lem:ddcond}, it follows that
        \begin{align*}
            \cV^{\pi}
            = \,  & \bE\bigg[\int_0^T \int_\bR \Big( \big( \mu(t) - \bE \left[ R(t) |\cH_{(t-d)^+}\right] \big) \pi (t, g) \, \bE \left[ \delta_G(g)|\cF_{(t-d)^+}\right]  \\
            & + \bE \left[ R(t) |\cH_{(t-d)^+}\right] - \frac{1}{2} \, \sigma^2(t) \, \pi^2 (t, g) \, \bE \left[ \delta_G(g)|\cF_{(t-d)^+}\right]                                                          \\
            & + \sigma(t) \, \pi (t, g) \, \bE \left[ D_{t^+} \, \delta_G(g)  |\cF_{(t-d)^+}\right] \Big) \, dg \, dt \bigg].
        \end{align*}
        Then, for all $t \in [0,T]$ and $g \in \supp(G)$, the optimal portfolio $\hat{\pi}(t,g)$ is
        \begin{equation*}
            \hat{\pi}(t,g) = \frac{\mu(t) - \bE \left[ R(t) |\cH_{(t-d)^+}\right]}{\sigma^2(t)} + \frac{\alpha_d(t,g)}{\sigma(t)}.
        \end{equation*}
        Substituting $\hat{\pi}(t,g)$ into equation~\eqref{eq:vk10}, we have that
        \begin{equation*}
            \cV^{\hat{\pi}}
            = \int_0^T \bE\left[ \frac{\left(\mu(t) - \bE \left[R(t)|\cH_{(t-d)^+}\right]\right)^2}{2 \, \sigma^2(t)} + R(t) + D_{t^+} \left( \alpha_d(t,G) \right) - \frac{1}{2} \alpha^2_d(t,G) \right] dt.
        \end{equation*}
        After subtracting~\eqref{eq:V.vk} and applying some algebraic manipulations, the result follows.
    \end{proof}
    \begin{corollary}
        \label{cor:vk}
        In the model~\eqref{eq:vk9} with $G=B(T)$, the AIT does not invariably obtain more expected utility than the traditional trader.
        Ultimately, this depends on the parameters of the interest rate evolution process $R(t)$ and the volatility of the stock $\sigma(t)$, since the difference in the expected logarithm of the final wealth is
        \begin{equation*}
            \frac{d}{2 \, T} + \frac{1}{2} \ln \left( \frac{T}{d} \right) -  \frac{\xi^2}{4 \, a} \int_0^T \frac{1-e^{-2a(t\wedge d)}}{\sigma^2 (t)} \, dt \, < \, \infty.
        \end{equation*}
    \end{corollary}
    \begin{proof}
        Applying Lemma~3.1 of~\cite{dauria.salmeron21}, we have that, for $0 \le s \le t$,
        \begin{equation*}
            \left( R(t)\vert r(s) \right)
            \sim
            \Normal\left(b-(b-r(s)) e^{-a(t-s)},\ \xi^2\frac{1-e^{-2a(t-s)}}{2a}\right) \ ,
        \end{equation*}
        and therefore, with $x \wedge y = \min(x,y)$,
        \begin{equation*}
            \Var \left[R(t)|\cH_{(t-d)^+}\right] = \xi^2\frac{1-e^{-2a(t \wedge d)}}{2a}.
        \end{equation*}

        The result follows directly from Theorem~\ref{th:vk3} and Corollary~\ref{co:bs}, where the parameters $a$ and $\xi$ are non-negative constants, and $\sigma(t)$ is a deterministic continuous function that is bounded away from zero.
    \end{proof}

    The results of Corollary~\ref{cor:vk} suggest that the following definition makes sense.
    \begin{definition}
        \label{def:vk}
        The \emph{temporal value} of the information is defined as the value $d^* \in [0,T]$ such that
        \begin{equation}
            \label{eq:temp.value}
            \frac{d^*}{2 \, T} + \frac{1}{2} \ln \left( \frac{T}{d^*} \right) -  \frac{\xi^2}{4 \, a} \int_0^T \frac{1-e^{-2a(t\wedge d^*)}}{\sigma^2 (t)} \, dt = 0,
        \end{equation}
        or $d^*=\infty$ in case such value does not exist.
    \end{definition}

\begin{figure}%
        \centering
        \subfloat[\centering $\xi=1$]{{\includegraphics[height=4cm]{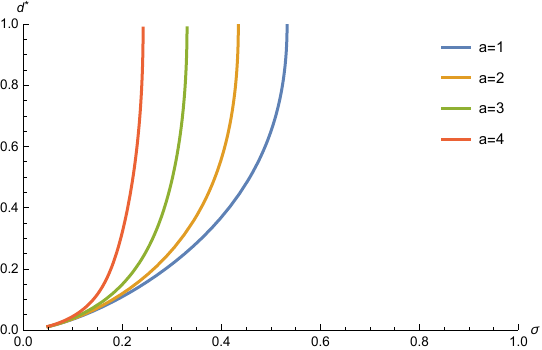} }}%
        \qquad
        \subfloat[\centering $a=1$]{{\includegraphics[height=4cm]{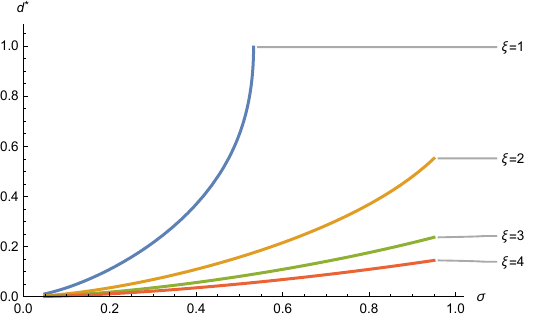} }}%
        \caption{Parametric representation of the temporal value of the information $d^*$ as a function of $a$ and $\xi$ for the Vasicek model with two delays and $G=B(T)$. $\sigma(t)$ is assumed to be constant equal to $\sigma$ and $T=1$.}%
        \label{fig:example}%
    \end{figure}

    Note that, since $d/(2T) + \ln \left( T/d \right)/2$ has no roots on $d \in [0,T]$, and since the integral is positive, $\xi \neq 0$ is a necessary condition for $d^*<\infty$.
    Figure~\ref{fig:example} plots the temporal value of the information as a function of the parameters $a$ and $\xi$, for $\sigma(t)=\sigma$ and $T=1$.

    \begin{remark}
    Other two well-known short-rate models were listed in Section~\ref{sec:fm}: the one-factor HW~\cite{hull, hullwhite} and the CIR~\cite{hull, lambertonlapeyre, cir} models. Under the hypotheses listed there, the main results for the Vasicek model, namely Theorems~\ref{th:vk1} and~\ref{th:vk3}, hold identically for them as well. This is because they are stochastic models for the short-rate that keep their first two moments finite; something that is shown along with explicit representation formulas for these moments in Section~\ref{sec:fm}. Note, however, that Corollary~\ref{cor:vk} and Definition~\ref{def:vk} do not follow identically for these two models. This is so since they depend on the explicit form of the conditioned variance of the short-rate, which differs from that of the Vasicek model. It is possible, nevertheless, to derive the analogous results for the HW and CIR models, however, they will yield more complicated expressions due to the more involved formulas for their variances.
    \end{remark}

\begin{remark}
A related problem to the one analyzed here is to assume that the AIT can choose portfolios from the class of $\bG$-predictable processes, where $\bG =\lbrace \cG_t \rbrace _{t\in [0,T]}$ is given by
    \begin{align*}
        \cG_t = \,  & \lbrace \sigma(B(s)): 0 \leq s \leq \left(t-d_1\right)^{+} \rbrace \vee \lbrace \sigma(W(s)): 0 \leq s \leq \left(t-d_2\right)^{+} \rbrace \vee \sigma \left(G \right) \\
        = \,  & \cF_{(t-d_1)^+} \vee \cH_{(t-d_2)^+} \vee \sigma \left(G \right),
    \end{align*}
with $d_1 \neq d_2$ ($d_1=d_2$ is the case already analyzed). Note that, in this generalized case, it is possible to repeat the proof of Theorem~\ref{th:vk3} to find the optimal portfolio
        \begin{equation*}
            \hat{\pi}(t,g)
            = \frac{\mu(t) - \bE\left[R(t)|\cH_{(t-d_2)^+}\right]}{\sigma^2(t)} + \frac{\alpha_{d_1}(t,g)}{\sigma(t)},
        \end{equation*}
        and the additional value
        \begin{equation*}
            \Delta\cV^{\hat{\pi}}
            = \int_0^T \bE\left[\frac{\Var \left[R(t)|\cH_{(t-d_2)^+}\right]}{2 \, \sigma^2(t)}
            + D_{t^+} \left( \alpha_{d_1}(t,G) \right) - \frac{1}{2} \alpha^2_{d_1}(t,G) \right] \, dt.
        \end{equation*}
        Nevertheless, we are not going to further discuss this case as, on one hand, we find it reasonable from a modelling viewpoint to assume the equality of both delays and, on the other hand, if these values were different, then it would not be possible to define a temporal value of the information as introduced in Definition~\ref{def:vk}.
\end{remark}

    \section{Conclusions}\label{sec:end}
    This work has been devoted to study the non-adapted version of the classical portfolio optimization problem in a financial market where a trader with privileged or insider information is present, but s/he receives the current market information with a delay (and hence the term AIT, i.~e. {\it asymmetrically informed trader}).
    This approach was carried out by O.~Draouil and B.~{\O}ksendal in~\cite{do2}, and our present aim has been to extend it for several models that are well-known in the financial mathematics literature. The results obtained for them has been compared to the performance of the traditional trader.

    Our main tools to study these financial problems have been, on one hand, the anticipating stochastic calculus (via posing them as Russo-Vallois forward stochastic differential equations) and, on the other hand, the white noise theory.
    In Section~\ref{sec:rvmc}, we have recalled some notions related to forward stochastic integration~\cite{russovallois} and Malliavin calculus~\cite{noep, do1}, which are necessary for the computation of the results, such as the Malliavin derivative and the Donsker delta function.

    Precisely, we have computed the optimal portfolio $\hat{\pi}$ and the expected logarithm of the final wealth (the expected logarithmic utility) for the AIT and the traditional trader. We have proved the superiority of the insider information despite the presence of the delay in a single information flow, for the Black-Scholes-Merton~\cite{bsm, hull, lambertonlapeyre}, the Heston~\cite{hull, heston}, and the Vasicek~\cite{hull, lambertonlapeyre, vasicek} models in Sections~\ref{sec:bsm},~\ref{sec:hes}, and~\ref{sec:vk} respectively.
    In essence, the optimal portfolios $\hat{\pi}(t)$ are given by
    \begin{equation*}
        \hat{\pi}(t,B(T)) = \frac{\mu(t) - \hat{\varrho}(t)}{\Sigma^2(t)} + \frac{\alpha_d(t,G)}{\Sigma(t)}
        \qquad \text{and} \qquad
        \hat{\pi}(t) = \frac{\mu(t) - \hat{\varrho}(t)}{\Sigma^2(t)},
    \end{equation*}
    for the AIT and the traditional trader respectively, where $\hat{\varrho}(t) \in \lbrace \rho(t), \rho(t), R(t) \rbrace$ and $\Sigma(t) \in \left\lbrace \sigma(t), \sqrt{V(t)}, \sigma(t) \right\rbrace$ denote the interest rate and the volatility of each of the models herein considered (in their order of appearance).
    In words, our results indicate that, under equivalent hypotheses, the same conclusions can be drawn for each model.
    The AIT always obtains more expected utility than the traditional trader despite the presence of a delay $d>0$ in the information flow of the stock.
    Whenever we assume the simpler information $G=B(T)$, the differences in the expected logarithms of the final wealth coincide in every case, and their common value is explicitly given by
    \begin{equation*}
        \frac{d}{2 \, T} + \frac{1}{2} \ln \left( \frac{T}{d} \right) \, \in \, (0,\infty) \qquad \text{for} \quad d \, \in \, (0,T].
    \end{equation*}
    
    Finally, in Section~\ref{sec:vk.delay}, we have analyzed the Vasicek model with delays present in two information flows: the stock and the interest rate time evolution. In such a case, we have concluded that, contrary to what happened in all of the other cases, the AIT does not necessarily obtain more utility than the traditional trader.
    This actually depends on the model parameters, as the difference in expected utilities can be reduced to the quadrature
    \begin{equation*}
        \frac{d}{2 \, T} + \frac{1}{2} \ln \left( \frac{T}{d} \right) -  \frac{\xi^2}{4 \, a} \int_0^T \frac{1-e^{-2a(t\wedge d)}}{\sigma^2 (t)} \, dt,
    \end{equation*}
    if we assume the simpler insider information $G=B(T)$. This result allowed us to introduce in Definition~\ref{def:vk} the (to the best of our knowledge) novel concept of {\it temporal value of the information}. With this new notion, the insider information can be valued, not just in terms of utility, as it is usually done, but also in terms of time. This means that the insider information can be measured as the temporal delay in following the market conditions, in the sense that future information can compensate such a delay.

    In summary, we have clarified certain instances of how an asymmetrically informed trader, one who has simultaneously privileged information about the future but follows the market conditions with a certain delay, performs in a financial market in comparison to a traditional trader. We have found that the privileged information always overcompensates the delay in a single information flow, but enters into a competition with it when two information flows are delayed. This, in turn, has permitted us to value privileged information in terms of time (what complements the traditional valuation in terms of utility). Probably, our results need to be reconfirmed with other models, particularly those closer to the financial practice. Presumably, such study would require an extensive numerical investigation that would complement the theoretical development herein presented.

    \section*{Acknowledgements}
    This work has been partially supported by the Government of Spain (Ministerio de Ciencia e Innovaci\'on) and the European Union through Projects PID2020-116694GB-I00, PID2021-125871NB-I00, TED2021-131844B-I00 /AEI /10.13039 /501100011033 /Uni\'on Europea Next Generation EU /PRTR, and CPP2021-008644 /AEI /10.13039 /501100011033 /Uni\'on Europea Next Generation EU/ PRTR.
    The first author is a member of the Gruppo Nazionale Calcolo Scientifico-Istituto Nazionale di Alta Matematica (GNCS-INdAM) and acknowledges also the partial support from the Italian SID project BIRD239937/23.

    \bibliographystyle{alpha}
    \bibliography{pofiiidbib}

\newcommand{\etalchar}[1]{$^{#1}$}
\begin{thebibliography}{DN{\O}P09}

\bibitem[Bac00]{bachelier}
L~Bachelier.
\newblock Th\'eorie de la sp\'eculation.
\newblock {\em Annales Scientifiques de l'Ecole Normale Sup\'erieure},
  17:21--86, 1900.

\bibitem[BE18]{bastons2018triple}
Joan~C Bastons and Carlos Escudero.
\newblock A triple comparison between anticipating stochastic integrals in
  financial modeling.
\newblock {\em Communications on Stochastic Analysis}, 12(1):73--87, 2018.

\bibitem[B{\O}05]{bo}
F~Biagini and B~{\O}ksendal.
\newblock A general stochastic calculus approach to insider trading.
\newblock {\em Applied Mathematics \& Optimization}, 52:167--181, 2005.

\bibitem[BS73]{bsm}
F~Black and M~Scholes.
\newblock The pricing of options and corporate liabilities.
\newblock {\em Journal of Political Economy}, 81(3):637--654, 1973.

\bibitem[CIR85]{cir}
JC~Cox, JE~Ingersoll, and SA~Ross.
\newblock A theory of the term structure of interest rates.
\newblock {\em Econometrica}, 53(2):385--408, 1985.

\bibitem[DN{\O}P09]{noep}
G~Di~Nunno, B~{\O}ksendal, and F~Proske.
\newblock {\em Malliavin Calculus for L\'evy Processes with Applications to
  Finance}.
\newblock Springer, Berlin, 2009.

\bibitem[D{\O}15]{do1}
O~Draouil and B~{\O}ksendal.
\newblock A {D}onsker delta functional approach to optimal insider control and
  applications to finance.
\newblock {\em Communications in Mathematics and Statistics}, 3:365--421, 2015.

\bibitem[D{\O}19]{do2}
O~Draouil and B~{\O}ksendal.
\newblock A white noise approach to optimal insider control of systems with
  delay.
\newblock {\em Journal of Mathematical Analysis and Applications},
  476:101--119, 2019.

\bibitem[DS24]{dauria.salmeron21}
Bernardo D'Auria and Jos{\'e}~Antonio Salmer{\'o}n.
\newblock Optimal portfolios with anticipating information on the stochastic
  interest rate.
\newblock {\em Decisions in Economics and Finance}, 2024.
\newblock doi:10.1007/s10203-024-00463-z.

\bibitem[EE22]{elizalde2022chances}
Mauricio Elizalde and Carlos Escudero.
\newblock Chances for the honest in honest versus insider trading.
\newblock {\em SIAM Journal on Financial Mathematics}, 13(2):SC39--SC52, 2022.

\bibitem[EEI22]{elizalde2022apo}
Mauricio Elizalde, Carlos Escudero, and Tomoyuki Ichiba.
\newblock Optimal investment with insider information using skorokhod \&
  russo-vallois integration.
\newblock arXiv:2211.07471, 2022.

\bibitem[Esc18]{escudero2018}
Carlos Escudero.
\newblock A simple comparison between {S}korokhod \& {R}usso-{V}allois
  integration for insider trading.
\newblock {\em Stochastic Analysis and Applications}, 36(3):485--494, 2018.

\bibitem[GHL{\etalchar{+}}93]{oksendal2}
H.~Gjessing, H.~Holden, T.~Lindstr{\o}m, B.~{\O}ksendal, J.~Ub{\o}e, and T.-S.
  Zhang.
\newblock The wick product.
\newblock In H.~Niemi, G.~H{\"o}gnas, A.~N. Shiryaev, and A.~V. Melnikov,
  editors, {\em Vol. 1 Proceedings of the Third Finnish-Soviet Symposium on
  Probability Theory and Mathematical Statistics, Turku, Finland, August 13-16,
  1991}, pages 29--67. De Gruyter, Berlin, Boston, 1993.

\bibitem[Hes93]{heston}
{\relax S-L}~Heston.
\newblock A closed-form solution for options with stochastic volatility with
  applications to bond and currency options.
\newblock {\em The Review of Financial Studies}, 6(2):327--343, 1993.

\bibitem[H{\O}UZ10]{HOUZ10}
Helge Holden, Bernt {\O}ksendal, Jan Ub{\o}e, and Tusheng Zhang.
\newblock {\em Stochastic Partial Differential Equations: A Modeling, White
  Noise Functional Approach}.
\newblock Springer New York, 2010.

\bibitem[Hul03]{hull}
JC~Hull.
\newblock {\em Options, Futures and Other Derivatives}.
\newblock Prentice Hall, 2003.

\bibitem[HW90]{hullwhite}
JC~Hull and A~White.
\newblock Pricing interest-rate-derivative securities.
\newblock {\em Review of Financial Studies}, 3(4):573--592, 1990.

\bibitem[It{\^o}44]{ito1}
K~It{\^o}.
\newblock Stochastic integral.
\newblock {\em Proceedings of the Imperial Academy}, 20(8):519--524, 1944.

\bibitem[It{\^o}46]{ito2}
K~It{\^o}.
\newblock On a stochastic integral equation.
\newblock {\em Proceedings of the Imperial Academy}, 22(2):32--35, 1946.

\bibitem[It{\^o}51]{ito3}
K~It{\^o}.
\newblock {Multiple Wiener Integral}.
\newblock {\em Journal of the Mathematical Society of Japan}, 3(1):157 -- 169,
  1951.

\bibitem[JYC09]{jeanblanc2009}
Monique Jeanblanc, Marc Yor, and Marc Chesney.
\newblock {\em Mathematical methods for financial markets}.
\newblock Springer Science \& Business Media, 2009.

\bibitem[KLS87]{karatzas1}
I~Karatzas, JP~Lehoczky, and SE~Shreve.
\newblock Optimal portfolio and consumption decisions for a ``small investor''
  on a finite horizon.
\newblock {\em SIAM Journal on Control and Optimization}, 25(6):1557--1586,
  1987.

\bibitem[Kuo06]{kuo}
{\relax H-H}~Kuo.
\newblock {\em Introduction to Stochastic Integration}.
\newblock Springer, New York, 2006.

\bibitem[LL96]{lambertonlapeyre}
D~Lamberton and B~Lapeyre.
\newblock {\em Introduction to Stochastic Calculus Applied to Finance}.
\newblock Chapman \& Hall, 1996.

\bibitem[LNN03]{leon}
JA~Le\'on, R~Navarro, and D~Nualart.
\newblock An anticipating calculus approach to the utility maximization of an
  insider.
\newblock {\em Mathematical Finance}, 13(1):171--185, 2003.

\bibitem[Mar52]{markowitz}
H~Markowitz.
\newblock Portfolio selection.
\newblock {\em The Journal of Finance}, 7(1):77--91, 1952.

\bibitem[Mer69]{merton1}
RC~Merton.
\newblock Lifetime portfolio selection under uncertainty: The continuous-time
  case.
\newblock {\em The Review of Economics and Statistics}, 51(3):247--257, 1969.

\bibitem[Mer73]{merton2}
RC~Merton.
\newblock Theory of rational option pricing.
\newblock {\em Bell Journal of Economics and Management Science},
  4(1):141--183, 1973.

\bibitem[{\O}ER17]{oksendal2017}
Bernt {\O}ksendal and Elin Engen~R{\o}se.
\newblock A white noise approach to insider trading.
\newblock In {\em Let Us Use White Noise}, pages 191--203. World Scientific,
  2017.

\bibitem[{\O}ks03]{oksendal1}
B~{\O}ksendal.
\newblock {\em Stochastic Differnetial Equations: An Introduction with
  Applications}.
\newblock Springer, Berlin, 2003.

\bibitem[PK96]{pk}
I~Pikovsky and I~Karatzas.
\newblock Anticipative portfolio optimization.
\newblock {\em Advances in Applied Probability}, 28(4):1095--1122, 1996.

\bibitem[RV93]{russovallois}
F~Russo and P~Vallois.
\newblock Forward, backward and symmetric stochastic integration.
\newblock {\em Probability Theory and Related Fields}, 97:403--421, 1993.

\bibitem[Vas77]{vasicek}
O~Vasicek.
\newblock An equilibrium characterization of the term structure.
\newblock {\em Journal of Financial Economics}, 5:177--188, 1977.

\bibitem[YW71]{watyam}
T~Yamada and S~Watanabe.
\newblock On the uniqueness of solutions of stochastic differential equations.
\newblock {\em Journal of Mathematics of Kyoto University}, 11(1):155--167,
  1971.

\end{thebibliography}

\end{document}